\documentclass[aps,prd,preprint,groupedaddress,longbibliography,nofootinbib]{revtex4-1}

\usepackage{amsmath}
\usepackage{xcolor}
\newcommand{\Sb}{S_{\mathrm{bare}}}
\newcommand{\Thetab}{\Theta^\mathrm{bare}}
\newcommand{\RL}{\mathcal{R}_\Lambda}
\newcommand{\nt}{\notag}

\newcommand{\lb}{\left\lbrace}
  \newcommand{\rb}{\right\rbrace}
\newcommand{\Op}{\mathcal{O}}
\newcommand{\N}{\mathcal{N}}
\newcommand{\G}{\mathcal{G}}

\newcommand{\eff}{\mathrm{eff}}
\newcommand{\vev}[1]{\left\langle #1 \right\rangle}
\newcommand{\ep}{\epsilon}
\allowdisplaybreaks

\begin{document}


\title{Explicit construction of the energy-momentum tensor\\ in the
  large $N$ limit}


\author{Carlo Pagani} \email{carlo.pagani@guest.ung.si}
\affiliation{Institute f\"{u}r Physik (WA THEP)
  Johannes-Gutenberg-Universit\"{a}t\\ Staudingerweg 7, 55099 Mainz,
  Germany}

\author{Hidenori Sonoda} \email[Visiting Research Associate
till 29 September 2026,~]{h-sonoda@pobox.com} \affiliation{Department of
  Physics and Astronomy, The University of Iowa, Iowa City, Iowa
  52242, USA}



\date{\today}

\begin{abstract}
  We construct the energy-momentum tensor of the O($N$) linear sigma
  model explicitly in the large $N$ limit using the exact
  renormalization group (ERG) formalism.  The energy-momentum tensor
  is obtained as a cutoff dependent functional of $N$ scalar field
  variables.  Our guiding principles behind the construction are
  twofold: first the energy-momentum tensor must satisfy the Ward
  identity for translation and rotation invariance, and second the
  energy-momentum tensor must satisfy a variant of the exact
  renormalization group equation.  In the limit that the momentum
  cutoff goes to zero, our energy-momentum tensor gives the
  one-particle irreducible (1PI) effective action with the insertion
  of a single energy-momentum tensor operator.  We verify that the
  energy-momentum tensor constructed satisfies the expected trace
  formula, and that the trace vanishes at the Wilson-Fisher critical
  point.
\end{abstract}


\maketitle

\section{Introduction\label{sec-introduction}}

The energy-momentum (EM) tensor encodes the conservation of energy,
momentum, and angular momentum. As such, it is a key quantity in
quantum field theory.
As proven by Belinfante and Rosenfeld \cite{BELINFANTE1940449,Rosenfeld1940},
the EM tensor can be made symmetric in the canonical formalism. 
The symmetry becomes manifest if one defines the EM tensor as a
variation of the matter action with respect to an external background
metric; the so defined EM tensor plays a fundamental role for the
equation of motion of gravitational theories.
The EM tensor also plays a key role in the construction of conformal
field theories \cite{DiFrancesco:1997nk}.
The issues related to renormalizability have been discussed in
\cite{Callan:1970ze} within perturbation theory.
%

In this paper we construct explicitly the EM tensor of the O$(N)$
linear sigma model in the large $N$ limit within the framework of the
exact renormalization group (ERG), also known as the functional
renormalization group (fRG).  In this framework we work with a Wilson
action that is a functional of field variables and depends on a
momentum cutoff.  Any local composite operator, regarded as an
infinitesimal change of the Wilson action, is also realized as a
functional dependent on the same cutoff.  A variant of the Wilson
action, namely the one-particle irreducible (1PI) part of the Wilson
action, is particularly convenient since it becomes the effective
action in the limit of the vanishing cutoff.  Correspondingly, the
cutoff dependent composite operator reduces to the effective action
with a single insertion of the composite operator.

By definition the EM tensor must satisfy the Ward identities
associated with translational and rotational invariance
\cite{Sonoda:2015pva}.
One of the advantages of working within the ERG framework is that we
only deal with finite quantities at all steps thanks to the presence
of a finite ultraviolet (UV) cutoff.
The relevant Ward identities are first introduced in terms of Wilson
actions and eventually translated to the associated
one-particle-irreducible (1PI) actions, or effective average
actions. (We refer the reader to
\cite{Berges:2000ew,Pawlowski:2005xe,Delamotte:2007pf,Igarashi:2009tj,Dupuis:2020fhh}
for a review of the ERG and the associated functionals.)

We focus on the the large $N$ limit of the O$(N)$ linear sigma model
in $D$ dimensions where $2 < D < 4$.  This allows us to explore a
genuinely non-perturbative, albeit tractable, example.  Our
construction of the large $N$ limit is particularly suitable to
describe the theory near criticality, i.e., the Wilson-Fisher fixed
point.
We work in the flat Euclidean space.  See Ref.~\cite{Heller:2021wan}
for an approach, in the context of ERG, to calculate the EM tensor in
spaces endowed with a metric of a different signature.
Thanks to our explicit construction we can also check conformal
invariance at criticality by calculating the trace of the EM tensor,
whose vanishing amounts to conformal invariance
\cite{Polchinski:1987dy}.  This statement holds true also within the
ERG framework
\cite{Rosten:2014oja,Delamotte:2015aaa,Sonoda:2015pva,Sonoda:2017zgl,Delamotte:2024xhn,Cabrera:2024rgy}.
Thus, our findings confirm that the ERG is a powerful framework to
discuss the realization of symmetries, despite the explicit presence
of a UV cutoff, see
\cite{Delamotte:2015aaa,Rosten:2014oja,Rosten:2016zap,Rosten:2016nmc,Sonoda:2017zgl,DePolsi:2019owi}
for further discussions concerning the ERG and conformal invariance.

The paper is organized as follows.
Sec.~\ref{sec-review} is dedicated to a somewhat lengthy review of the
ERG formalism, the EM tensor via the relevant Ward identities, and the
large $N$ limit in the ERG formalism.  We hope this section makes the
paper self-contained.
The EM tensor is constructed explicitly in Sec.~\ref{sec-EM} by
imposing the required Ward identities.
As we shall see, this determines the EM tensor up to two coefficients,
which are eventually fixed by requiring that the EM tensor satisfy the
required ERG differential equation.  Within the ERG framework,
composite operators are introduced as infinitesimal changes of the
Wilson action, and their cutoff dependence is given by the ERG
differential equation.  We explain all this in Sec.~\ref{sec-comp}.
Finally, in Sec.~\ref{sec-completion}, we complete the construction of
the EM tensor by fixing the two constants and compute the trace.
This is followed by a short Sec.~\ref{sec-zerolimit} where we take the
cutoff to zero to obtain the effective action with the insertion of a
single EM tensor.
In Sec.~\ref{sec-conclusions} we summarize our findings and discuss
possible developments.
Some technicalities are clarified in two Appendices.

We use the shorthand notation
\[
  \int_x = \int d^D x,\quad\delta (x) = \delta^{(D)} (x)
\]
for the integrals in coordinate space, and
\[
\int_p = \int \frac{d^D p}{(2 \pi)^D},\quad \delta (p) = (2 \pi)^D
  \delta^{(D)} (p)
\]
for those in the momentum space.  We denote the Fourier transform of a
function $f(x)$ in the coordinate space by $\tilde{f} (p)$ in the
momentum space.

\section{Review of the energy-momentum tensor and large
  $N$\label{sec-review}}

In this section we review the energy-momentum tensor and the large $N$
limit within the ERG formalism.  The energy-momentum tensor of a
scalar field theory has been discussed in details within the context
of a Wilson action and ERG in \cite{Sonoda:2015pva}.  In order to make
this paper self-contained as much as possible, we introduce a little
more pedagogical approach.  As for the large $N$ limit within the ERG
formalism, we would like to follow a particular method given in
\cite{Sonoda:2023ohb}.

\subsection{Quick review of ERG}

Let $\Sb [\phi]$ be the bare action of a renormalized theory of the
scalar field $\phi$ in $D$-dimensional Euclidean space.  We define a
functional $W_\Lambda [J]$ by the functional integral
\begin{equation}
  e^{W_\Lambda [J]}
 \equiv \int [d\phi] \exp \left[ \Sb [\phi] - \frac{1}{2} \int_{x,y}
 \RL (x-y) \phi (x) \phi (y) + \int_x  J(x) \phi
  (x) \right]\,,\label{sec2-WL}
\end{equation}
where $\RL (x)$ is a positive cutoff function that has a range of
$\frac{1}{\Lambda}$, and normalized by
\begin{equation}
  \int_x\, \RL (x) = \Lambda^2\,.
\end{equation}
We take $\RL (x)$ to be rotation invariant, dependent only on $x^2$.
The Fourier transform of $\RL (x)$, defined by
\begin{equation}
  R_\Lambda (p) \equiv \int_x\, e^{- i p x} \RL (x)\,,
\end{equation}
can be considered as a momentum dependent squared mass, suppressing
the fluctuations of the fields with momenta less than $\Lambda$.  We
assume that $R_\Lambda (p)$ has the $\Lambda$ dependence given by
\begin{equation}
  R_\Lambda (p) = \Lambda^2 R (p/\Lambda)\,.
\end{equation}
For example, $R (p/\Lambda) = e^{- \frac{p^2}{\Lambda^2}}$ gives
\begin{equation}
  \RL (x) = \int_p e^{i p x} R_\Lambda (p) 
  = \Lambda^2 e^{- \frac{1}{4} \Lambda^2 x^2} \left(\frac{\Lambda^2}{4
      \pi}\right)^{\frac{D}{2}}\,. 
\end{equation}
Hence, as far as $W_\Lambda [J]$ is concerned, we can interpret $\Lambda$
as an infrared (IR) cutoff.  Since
\begin{equation}
  \lim_{\Lambda \to 0+} \RL (x) = 0\,,
\end{equation}
we find
\begin{equation}
  \lim_{\Lambda \to 0+} W_\Lambda [J] = \mathcal{W} [J]\,,
\end{equation}
where $\mathcal{W} [J]$ is the generating functional of the
connected correlation functions, defined by
\begin{equation}
  e^{\mathcal{W} [J]} \equiv \int [d\phi] \exp \left( \Sb [\phi] +
    \int_x\, J(x) \phi (x) \right)\,.
\end{equation}

We then introduce the one-particle-irreducible (1PI) Wilson action (or
effective average action) $\Gamma_\Lambda [\Phi]$ as the Legendre
transform of $W_\Lambda [J]$:
\begin{equation}
  \Gamma_\Lambda [\Phi] - \frac{1}{2}
  \int_{x,y}\, \RL (x-y) \Phi (x) \Phi (y)
  \equiv W_\Lambda [J] - \int_x\, J(x) \Phi  (x)\,,
\end{equation}
where
\begin{equation}
  \Phi (x) \equiv \frac{\delta W_\Lambda [J]}{\delta J(x)}\,.
\end{equation}
In the limit $\Lambda \to 0+$, $\RL$ vanishes, and we obtain the
effective action
\begin{equation}
  \lim_{\Lambda \to 0+} \Gamma_\Lambda [\Phi] = \Gamma_\eff [\Phi]
\end{equation}
as the Legendre transform of $\mathcal{W} [J]$.

The cutoff dependence of $W_\Lambda [J]$ is given by the ERG
differential equation
\begin{equation}
  - \Lambda \partial_\Lambda W_\Lambda [J]
  = \frac{1}{2} \int_{x,y}\, \Lambda \partial_\Lambda \RL
  (x-y)\, \left( \frac{\delta W_\Lambda
      [J]}{\delta J(x)} \frac{\delta W_\Lambda [J]}{\delta J(y)} +
    \frac{\delta^2 W_\Lambda [J]}{\delta J(x) \delta J(y)} \right)\,.
\end{equation}
That of $\Gamma_\Lambda [\Phi]$ is given by
\begin{equation}
  - \Lambda \partial_\Lambda \Gamma_\Lambda [\Phi]
  = \frac{1}{2} \int_{x,y}\,
  \Lambda \partial_\Lambda \RL (x-y)\,  \G_\Lambda (x,y) [\Phi]\,,
  \label{sec2-ERG-full}
\end{equation}
where
\begin{equation}
  \G_\Lambda (x,y) [\Phi] \equiv \frac{\delta^2 W_\Lambda [J]}{\delta J(x)
    \delta J(y)} = \frac{\delta \Phi (x)}{\delta J(y)} = \frac{\delta
    \Phi (y)}{\delta J(x)}
\end{equation}
is regarded as a functional of $\Phi$.  Since the inverse Legendre
transformation gives
\begin{equation}
  J(x) = \int_y\, \RL (x-y) \Phi (y) - \frac{\delta \Gamma_\Lambda
    [\Phi]}{\delta \Phi (x)}\,,\label{sec2-JPhi}
\end{equation}
we obtain
\begin{equation}
  \frac{\delta J(x)}{\delta \Phi (y)}
  = \RL (x-y) - \frac{\delta^2 \Gamma_\Lambda [\Phi]}{\delta \Phi (x)
    \delta \Phi (y)}\,.
\end{equation}
Hence, $\G_\Lambda = \frac{\delta \Phi}{\delta J}$ is obtained as the inverse of
$\frac{\delta J}{\delta \Phi}$:
\begin{equation}
  \int_y\, \G_\Lambda (x,y) [\Phi] \left(
    \RL (y-z) - \frac{\delta^2 \Gamma_\Lambda [\Phi]}{\delta \Phi (y)
      \delta \Phi (z)}\right) = \delta (x-z)\,.\label{sec2-GL}
\end{equation}

\subsection{Quick review of the energy-momentum tensor and its Ward identity}

We assume that the bare action is invariant under translations in
space.  Under the infinitesimal change of field variables
\begin{equation}
  \phi (x) \longrightarrow \phi (x) + \ep_\mu (x) \partial_\mu \phi
  (x)\,,\label{sec2-changephi}
\end{equation}
the action changes as
\begin{equation}
  \delta \Sb [\phi] = \int_x\, \ep_\mu (x) \partial_\mu \phi (x)
  \frac{\delta \Sb [\phi]}{\delta \phi (x)}\,.
\end{equation}
Since this should vanish for constant $\ep_\mu$, we should be able to
write this as
\begin{equation}
  \delta \Sb [\phi] = \int_x\, \ep_\mu  (x) \partial_\nu
  \Thetab_{\nu\mu} (x)\,.
  \label{sec2-deltaSbare}
\end{equation}
Since the functional integral (\ref{sec2-WL}) does not change under
(\ref{sec2-changephi}), we obtain
\begin{align}
&  \int [d\phi] \int_x\, \ep_\mu (x) \left( \partial_\nu
  \Thetab_{\nu\mu} (x) - \int_y\, \RL (x-y) \partial_\mu \phi (x)
  \phi (y) + J(x) \partial_\mu \phi (x) \right)\nt\\ 
&\quad \times
  \exp \left[ \Sb [\phi]
    - \frac{1}{2} \int_{x,y}\,\RL (x-y) \phi (x) \phi (y) + \int_x
  \, J(x) \phi (x) \right] = 0\,.\label{sec2-Ward-bare}
\end{align}
As a functional of $J$, we define the energy-momentum (EM) tensor
$\Theta_{\mu\nu} (x)$ by
\begin{align}
  \Theta_{\mu\nu} (x)\,e^{W_\Lambda [J]}
  &\equiv \int [d\phi]\, \Thetab_{\mu\nu} (x) \exp \Bigg[ \Sb [\phi] \nt\\
  &\quad - \frac{1}{2} \int_{x,y}\, \RL (x-y) \phi (x) \phi (y)
    + \int_x\, J(x) \phi (x)\Bigg]\,.
\end{align}
We can rewrite (\ref{sec2-Ward-bare}) as the Ward identity satisfied
by the EM tensor:
\begin{equation}
  \partial_\nu \Theta_{\nu\mu} (x)\, e^{W_\Lambda [J]}
  = \left( \int_y\, \RL (x-y) \frac{\partial}{\partial x_\mu}
  \frac{\delta^2}{\delta J(x) \delta J(y)} - J(x) \frac{\delta}{\delta
    J(x)} \right) e^{W_\Lambda [J]}\,.
\end{equation}
This gives
\begin{equation}
  \partial_\nu \Theta_{\nu\mu} (x)
  = \int_y\, \RL (x-y) \lb \frac{\partial}{\partial x_\mu}
  \frac{\delta^2 W_\Lambda [J]}{\delta J(x) \delta J(y)} +
  \frac{\delta W_\Lambda [J]}{\delta J(y)} \partial_\mu \frac{\delta
      W_\Lambda [J]}{\delta J(x)} \rb - J(x) \partial_\mu \frac{\delta
      W_\Lambda [J]}{\delta J(x)}\,.
\end{equation}
Regarding $\Theta_{\mu\nu} (x)$ as a functional of $\Phi$ instead of
$J$, we can rewrite this further as
\begin{equation}
  \partial_\nu \Theta_{\nu\mu} (x)
  = \int_y\, \RL (x-y) \frac{\partial}{\partial x_\mu}
  \G_\Lambda (x,y) [\Phi]  + \frac{\delta \Gamma_\Lambda
    [\Phi]}{\delta \Phi (x)} \partial_\mu \Phi (x) \,,
  \label{sec2-Ward}
\end{equation}
where we have used (\ref{sec2-JPhi}).

So far we have only discussed the consequences of the translation
invariance of the theory.  We assume that the theory be invariant also
under rotations.  An infinitesimal rotation corresponds to
\begin{equation}
  \ep_\mu (x) = \ep_{\mu\nu} x_\nu\,,
\end{equation}
where $\ep_{\mu\nu} = - \ep_{\nu\mu}$ is an arbitrary constant
antisymmetric tensor.  Substituting this $\ep_\mu (x)$ into
(\ref{sec2-deltaSbare}), we obtain
\begin{equation}
\int_x\, \left(  \Thetab_{\mu\nu} (x) - \Thetab_{\nu\mu}
  (x)\right) = 0\,.
\end{equation}
Using a well known construction due to Belinfante
\cite{BELINFANTE1940449}and Rosenfeld \cite{Rosenfeld1940} (the
details are found in \cite{Sonoda:2015pva} for example), we can
redefine $\Theta_{\mu\nu} (x)$, while keeping the Ward identity
(\ref{sec2-Ward}) intact, so that the symmetry holds locally at any
$x$:
\begin{equation}
  \Theta_{\mu\nu} (x) = \Theta_{\nu\mu} (x)\,.\label{sec2-EM-symmetry}
\end{equation}
Hence, as a consequence of the invariance of the theory under both
translations and rotations, we can conclude the existence of a
symmetric EM tensor $\Theta_{\mu\nu} (x)$ that satisfies the Ward
identity (\ref{sec2-Ward}).

Note that $\Theta_{\mu\nu} (x)$ is a functional of $\Phi$, and it
should be more properly written as $\Theta_{\mu\nu} (x) [\Phi]$.  In
the limit $\Lambda \to 0+$, it becomes the effective action with a
single insertion of the energy-momentum tensor operator.  To be more
specific, we obtain
\begin{equation}
  \lim_{\Lambda \to 0+} \Theta_{\mu\nu} (x) [\Phi]
  = \sum_{n=0}^\infty \frac{1}{n!}
  \int_{x_1\cdots x_n}\, \Phi (x_1) \cdots \Phi (x_n)
  \, \vev{\Theta_{\mu\nu} (x)\, \phi (x_1) \cdots \phi
    (x_n)}^{\mathrm{1PI}}\,,
\end{equation}
where the suffix 1PI denotes the 1PI part of the correlation function
\begin{equation}
  \vev{\Theta_{\mu\nu} (x)\, \phi (x_1) \cdots \phi
    (x_n)} \equiv
  \int [d\phi] \, \Thetab_{\mu\nu} (x)\, \phi (x_1) \cdots \phi (x_n) \,
  e^{\Sb [\phi]}\,.
\end{equation}

\subsection{Quick review of the large $N$ limit}

In the O($N$) linear sigma model we have $N$ scalar fields
$\Phi^I\,(I=1,\cdots,N)$, and the Ward identity (\ref{sec2-Ward})
for the EM tensor is given by
\begin{equation}
  \partial_\mu \Theta_{\mu\nu} (x) = \int_y\, \RL (x-y)
  \frac{\partial}{\partial x_\nu} \frac{\delta \Phi^I (x)}{\partial
    J^I (y)} + \partial_\nu  \Phi^I (x) \frac{\delta \Gamma_\Lambda
    [\Phi]}{\delta \Phi^I (x)}\,,
  \label{sec2-Ward-Gamma-N}
\end{equation}
where the repeated index $I$ is summed from $1$ to $N$.  In the large
$N$ limit
\cite{Berlin-Kac:1952,Stanley:1968,Wilson:1972cf,Ma:1973zu,Schnitzer:1974ji,Schnitzer:1974ue,Coleman:1974jh,Moshe:2003xn},
the 1PI Wilson action is simplified as
\begin{equation}
  \Gamma_\Lambda [\Phi] = - \frac{1}{2} \int_x\, \partial_\mu
  \Phi^I (x) \partial_\mu \Phi^I (x) + N \Gamma_{I\Lambda} [\rho]\,,
\end{equation}
where
\begin{equation}
  \rho (x) \equiv  \frac{1}{2N} \Phi^I (x) \Phi^I (x) 
\end{equation}
is the O($N$) invariant squared scalar field of the mass dimension
$D-2$.  (For the large $N$ approximation within the ERG formalism, see
\cite{DAttanasio:1997yph,Morris:1997xj,Blaizot:2005xy,Litim:2018pxe}.
We are following \cite{Sonoda:2023ohb} for this part of review.)
Defining
\begin{equation}
  \sigma (x) \equiv \frac{\delta \Gamma_{I\Lambda} [\rho]}{\delta
    \rho (x)},\label{sec2-sigma}
\end{equation}
we obtain
\begin{equation}
  \frac{\delta \Phi^I (x)}{\delta J^J (y)}
  = \frac{\delta \Phi^J (y)}{\delta J^I (x)}
  \simeq \delta^{IJ} \G_{\Lambda} (x,y) [\sigma]
  + \cdots,\label{sec2-dPhidJ}
\end{equation}
where $\G_\Lambda (x,y) [\sigma]$ is given by
\begin{equation}
  \left( - \frac{\partial^2}{\partial y_\mu\partial
      y_\mu} - \sigma (y) \right) \G_{\Lambda} (x,y) [\sigma] +
  \int_z\, \G_\Lambda (x,z) [\sigma]  \RL (z-y)  = \delta (x-y)\,.
\label{sec2-GL-largeN}
\end{equation}
The suppressed terms in (\ref{sec2-dPhidJ}) do not give an order $N$
contribution to the trace over $I=J$.
Note that $\sigma (x)$ has the mass dimension $2$.  We now define the
high momentum propagator by
\begin{equation}
  h_\Lambda (x) = \int_p e^{i p x} \tilde{h}_\Lambda (p) \equiv \int_p
  e^{i p x} \frac{1}{p^2 + R_\Lambda (p)} \label{sec2-def-hL}
\end{equation}
that satisfies
\begin{equation}
 - \partial^2 h_\Lambda (x-y) + \int_z\,  h_\Lambda (x-z) \RL (z-y)
 = \delta (x-y)\,.\label{sec2-diffeq-hL}
\end{equation}
Using this, we can solve for $\G_\Lambda (x,y)[\sigma]$ as a geometric
series:
\begin{align}
  \G_\Lambda (x,y) [\sigma]
  &= h_\Lambda (x-y) + \int_{z_1} \, h_\Lambda (x-z_1) \sigma (z_1)
    h_\Lambda (z_1-y)\nt\\
  &\quad + \int_{z_1 ,z_2}\, h_\Lambda (x-z_1) \sigma (z_1)
    h_\Lambda (z_1-z_2) \sigma (z_2) h_\Lambda (z_2 - y) + \cdots\,.
\end{align}
In large $N$ the ERG differential equation (\ref{sec2-ERG-full})
reduces to that of $\Gamma_{I\Lambda} [\rho]$, given by
\begin{equation}
  - \Lambda \partial_\Lambda \Gamma_{I\Lambda} [\rho]
  = \frac{1}{2} \int_{x,y} \Lambda \partial_\Lambda \RL (x-y) \cdot
  \G_\Lambda (x,y) [\sigma]\,.\label{sec2-ERG-GammaIL}
\end{equation}

Using  (\ref{sec2-dPhidJ})  and
\begin{align}
  \frac{\delta \Gamma_\Lambda [\Phi]}{\delta \Phi^I (x)}
  &= \partial^2 \Phi^I (x) + N \frac{\delta \Gamma_{I\Lambda}
    [\rho]}{\delta \Phi^I (x)}\nt\\
  &= \partial^2 \Phi^I (x) + \sigma (x)  \Phi^I (x)\,,
\end{align}
we can rewrite  (\ref{sec2-Ward-Gamma-N})  as
\begin{align}
  \partial_\mu \Theta_{\mu\nu} (x)
  &= \int_y\, \RL (x-y) N \frac{\partial}{\partial x_\nu}
    \G_\Lambda (x,y) [\sigma] + \partial_\nu \Phi^I (x)
    \left( \partial^2 \Phi^I (x) + \sigma (x) \Phi^I (x) \right)\nt\\
  &= \partial_\nu \Phi^I (x) \cdot \partial^2 \Phi^I (x)
    + N \lb \int_y\, \RL (x-y) \frac{\partial}{\partial x_\nu}
    \G_\Lambda (x,y) [\sigma] + \partial_\nu \rho (x) \cdot \sigma (x)
    \rb\,.
    \label{sec2-Ward-to-solve}
\end{align}
The goal of this paper is to construct $\Theta_{\mu\nu} (x)$ as a
functional of $\Phi^I$ that satisfies this identity.

We close this section by giving explicitly the interaction part
$\Gamma_{I\Lambda} [\rho]$ of the 1PI Wilson action.  To be more
accurate we give its Legendre transform explicitly.  We define a
functional of $\sigma$ by
\begin{equation}
  F_\Lambda [\sigma] \equiv \Gamma_{I\Lambda} [\rho] - \int_x\,
  \sigma (x) \rho (x)\,.
\end{equation}
Because of (\ref{sec2-sigma}), this is a Legendre transformation.  The
ERG differential equation (\ref{sec2-ERG-GammaIL}) gives that of
$F_\Lambda [\sigma]$ as
\begin{equation}
  - \Lambda \partial_\Lambda F_\Lambda [\sigma]
  = \frac{1}{2} \int_{x,y}\, \Lambda \partial_\Lambda \RL (x-y) \cdot
  \G_{\Lambda} (x,y) [\sigma]\,.\label{sec2-ERG-FL}
\end{equation}
The most general solution is
\begin{equation}
  F_\Lambda [\sigma] = I_\Lambda [\sigma] + \bar{F} [\sigma],
\end{equation}
where $\bar{F} [\sigma]$ is an arbitrary functional independent of
$\Lambda$.  $I_\Lambda [\sigma]$ is a particular solution to
(\ref{sec2-ERG-FL}) given by
\begin{equation}
  I_\Lambda [\sigma]
  = \int_x \left( c_\Lambda  + c_{1\Lambda} \sigma (x) \right)
+ \frac{1}{2} \sum_{n=2}^\infty \frac{1}{n} \int_{x_1, \cdots, x_n}\,
    \sigma (x_1) \cdots \sigma (x_n) \, I_{n\Lambda} (x_1, \cdots,
    x_n)\,,\label{sec2-def-ILambda} 
  \end{equation}
  where
\begin{subequations}
\begin{align}
  c_\Lambda
  &\equiv   \frac{1}{2} \int_q \ln \left( q^2 \tilde{h}_\Lambda (q)\right)\,,\\
  c_{1\Lambda}
  &\equiv \frac{1}{2} \int_q \left( \tilde{h}_\Lambda
    (q)-\frac{1}{q^2}\right)\,,\label{sec2-c1Lambda}\\ 
  I_{n\Lambda} (x_1, \cdots, x_n)
  &\equiv h_\Lambda (x_1-x_2) 
    h_\Lambda (x_2-x_3) \cdots h_\Lambda (x_n - x_1)\,.\quad(n \ge 2)
\end{align}
\end{subequations}
Note that both $c_\Lambda$ and $c_{1\Lambda}$ are well defined, free
from UV divergences for $2 < D < 4$; they satisfy
\begin{subequations}
\begin{align}
  - \Lambda \partial_\Lambda c_\Lambda
  &= \frac{1}{2}
    \int_q \Lambda\partial_\Lambda R_\Lambda (q) \cdot
    \tilde{h}_\Lambda (q) = \frac{1}{2} \int_x \Lambda
    \partial_\Lambda \RL (x) \cdot h_\Lambda (x)\,,\\
  - \Lambda \partial_\Lambda c_{1\Lambda}
  &= \frac{1}{2} \int_q \Lambda \partial_\Lambda R_\Lambda (q) \cdot
    \tilde{h}_\Lambda (q)^2
    = \frac{1}{2} \int_{x,y} \Lambda \partial_\Lambda \RL (x-y) \cdot
    h_\Lambda (x) h_\Lambda (y)\,.
\end{align}
In the momentum space we can write
\begin{align}
&  \int_{x_1,\cdots,x_n} \sigma (x_1) \cdots \sigma (x_n) I_{n\Lambda}
  (x_1, \cdots, x_n)
  = \int_{p_1, \cdots, p_n} \delta (p_1+\cdots+p_n) \tilde{\sigma}
    (p_1) \cdots \tilde{\sigma} (p_n)\nt\\
  &\quad \times \int_q \tilde{h}_\Lambda (q) \tilde{h}_\Lambda (q+p_1)
    \tilde{h}_\Lambda (q+p_1+p_2) \cdots \tilde{h}_\Lambda
    (q+p_1+\cdots p_{n-1})\,,\quad (n \ge 2)
\end{align}
\end{subequations}
which makes the UV finiteness manifest.

In this paper we take
\begin{equation}
  \bar{F} [\sigma] = \int_x \left( f_1 \sigma (x) + f_2 \frac{1}{2}
    \sigma (x)^2 \right)\,,
\end{equation}
where $f_1$ has the mass dimension $D-2$, and $f_2 \ge 0$ has the mass
dimension $D-4 < 0$.  Since the Legendre transform of
$\bar{F} [\sigma]$ is
\begin{align}
 \bar{\Gamma} [\rho]
  &\equiv \bar{F} [\sigma] + \int_x \sigma (x) \rho (x)\nt\\
  &=  \int_x \left( - \frac{f_1^2}{2 f_2}  - \frac{f_1}{f_2} \rho (x) -
    \frac{1}{2 f_2} \rho (x)^2\right)\,,
\end{align}
we see that $\frac{f_1}{f_2}$ is a squared mass parameter, and
$\frac{1}{f_2}$ is a quartic coupling.  Both $\frac{f_1}{f_2}$ and
$\frac{1}{f_2}$ are relevant parameters for the Gaussian theory
$\Gamma_{I\Lambda} = 0$.  For the critical theory at $f_1=f_2=0$,
however, only $f_1$ is relevant, and $f_2$ is irrelevant.  At $f_1=0$
the theory is massless irrespective of $f_2 \ge 0$, and it approaches
the theory $f_2 = 0$ (the Wilson-Fisher fixed point) at long distances
and the theory $f_2 = +\infty$ (the Gaussian fixed point or the
massless free theory) at short distances.
  
\section{Explicit construction of the energy-momentum
  tensor\label{sec-EM}}

Let us recapitulate our goal: we would like to construct a symmetric
tensor $\Theta_{\mu\nu} (x)$, a functional of $\Phi^I$, that satisfies
the Ward identity (\ref{sec2-Ward-to-solve}).  In this section we
construct $\Theta_{\mu\nu} (x)$ inductively, and in
Sect.~\ref{sec-completion} we complete the construction using the
notion of $\Lambda$-dependent composite operators reviewed in
Sect.~\ref{sec-comp}.

We start with the energy-momentum tensor of the massless Gaussian
theory (see \cite{Sonoda:2015pva} within the context of ERG):
\begin{equation}
  \Theta^G_{\mu\nu} (x) \equiv
  \partial_\mu \Phi^I (x) \partial_\nu \Phi^I (x) - \frac{1}{2}
  \delta_{\mu\nu} \partial_\alpha \Phi^I (x) \partial_\alpha \Phi^I (x)\,.
\end{equation}
This satisfies
\begin{equation}
\partial_\mu \Theta^G_{\mu\nu} (x) = \partial^2 \Phi^I (x) \,
\partial_\nu \Phi^I (x)\,.
\end{equation}
We then obtain, from (\ref{sec2-Ward-to-solve}),
\begin{equation}
\partial_\mu \left( \Theta_{\mu\nu} (x) - \Theta_{\mu\nu}^G (x)
\right)
= N \left( - \sigma (x) \partial_\nu \frac{\delta F_\Lambda
    [\sigma]}{\delta \sigma (x)} + \int_y \RL (x-y)
  \frac{\partial}{\partial x_\nu} \G_\Lambda (x,y) [\sigma] \right)\,,
\end{equation}
where we have used the inverse Legendre transformation
\begin{equation}
  \rho (x) = - \frac{\delta F_\Lambda [\sigma]}{\delta \sigma
    (x)}\,.\label{sec3-varphi-sigma}
\end{equation}

It is convenient to construct
$\Theta_{\mu\nu} (x) - \Theta^G_{\mu\nu} (x)$ as a functional of
$\sigma$.  We first consider the dependence of the EM tensor on
$f_1, f_2$.  $\G_\Lambda (x,y) [\sigma]$ is independent of
$f_1, f_2$; only $\bar{F} [\sigma]$ in $F_\Lambda [\sigma]$ depends on
$f_1, f_2$.  Since
\begin{equation}
 - \partial_\nu \frac{\delta \bar{F} [\sigma]}{\delta
  \sigma (x)} \cdot \sigma (x)
  = - \partial_\nu \left( f_1 + f_2 \sigma (x)\right) \cdot \sigma
    (x) = - \partial_\mu \left( \delta_{\mu\nu} \frac{1}{2} f_2 \sigma
    (x)^2 \right)\,,
\end{equation}
we get the $f_2$ dependent contribution
\begin{equation}
  - N \delta_{\mu\nu}  \frac{1}{2} f_2 \sigma (x)^2
\end{equation}
to $\Theta_{\mu\nu} (x) - \Theta^G_{\mu\nu} (x)$.

Let us denote the remaining contribution as
\begin{equation}
  \Theta'_{\mu\nu} (x) \equiv \Theta_{\mu\nu} (x) - \Theta^G_{\mu\nu}
  (x) +  N \delta_{\mu\nu}  \frac{f_2}{2} \sigma (x)^2\,.\label{sec3-EMprime}
\end{equation}
This is determined from
\begin{align}
  \frac{1}{N} \partial_\mu \Theta'_{\mu\nu} (x)
  &= \int_y \RL (x-y) \frac{\partial}{\partial x_\nu} \G_\Lambda (x,y) [\sigma]
    - \sigma (x) \partial_\nu \frac{\delta I_\Lambda [\sigma]}{\delta
    \sigma (x)}\nt\\
  &= \int_y \RL (x-y) \frac{\partial}{\partial x_\nu} \lb
    h_\Lambda (x-y) + \int_z h_\Lambda (x-z) \sigma (z) h_\Lambda
    (z-y)\nt\right.\\
  &\left.\quad + \int_{z_1, z_2} h_\Lambda (x-z_1) \sigma (z_1) h_\Lambda
    (z_1-z_2) \sigma (z_2) h_\Lambda (z_2-y) + \cdots \rb\nt\\
  &\quad - \sum_{n=2}^\infty \frac{1}{2n} \sigma (x)
    \frac{\partial}{\partial x_\nu}
    \int_{x_1,\cdots,x_n} \sum_{i=1}^n \delta (x-x_i)  \sigma (x_1) \cdots
    \widehat{\sigma (x_i)} \cdots \sigma (x_n)\,
    I_{n\Lambda} (x_1, \cdots, x_n)\nt\\
  &= \int_y \RL (x-y) \frac{\partial}{\partial x_\nu} \int_z h_\Lambda
    (x-z) \sigma (z) h_\Lambda (z-y)\nt\\
  &\quad + \int_y \RL (x-y) \frac{\partial}{\partial x_\nu}
    \int_{z_1,z_2} h_\Lambda (x-z_1) \sigma (z_1) h_\Lambda (z_1-z_2)
    \sigma (z_2) h_\Lambda (z_2-y)\nt\\
  &\quad - \frac{1}{2} \sigma (x) \frac{\partial}{\partial x_\nu} \int_{x_1} I_{2\Lambda}
    (x,x_1) \sigma (x_1)\nt\\
  &\quad + \int_y \RL (x-y) \frac{\partial}{\partial x_\nu} \int_{z_1, z_2, z_3}
    h_\Lambda (x-z_1) \sigma (z_1) h_\Lambda (z_1-z_2) \sigma (z_2)
    h_\Lambda (z_2-z_3) \nt\\
  &\qquad\qquad \times \sigma (z_3) h_\Lambda (z_3-y)\nt
    - \frac{1}{2} \sigma (x) \frac{\partial}{\partial x_\nu} \int_{x_1, x_2}
    I_{3\Lambda} (x,x_1,x_2) \sigma (x_1) \sigma (x_2)\nt\\
  &\quad + \textrm{terms of higher orders in $\sigma$'s}.
    \label{sec3-expansion}
\end{align}
We determine the terms of order $\sigma, \sigma^2$, and $\sigma^3$.
Then it is easy to guess the higher order terms.

\subsection{$\sigma$ term}

We find
\begin{equation}
 \left(\textrm{$\sigma$\,term of (\ref{sec3-expansion})}\right)
  = \int_y \RL (x-y) \frac{\partial}{\partial x_\nu} \int_z h_\Lambda
  (x-z) \sigma (z) h_\Lambda (z-y)\,.
\end{equation}
We define a scalar function $A_\Lambda (x)$ by
\begin{subequations}
  \label{sec3-def-A}
\begin{equation}
\partial_\nu A_\Lambda (x) \equiv  \partial_\nu h_\Lambda (x) \int_y
\RL (x-y) h_\Lambda (y)\label{sec3-def-A-a}
\end{equation}
and
\begin{equation}
\int_x A_\Lambda (x) = \textrm{finite}\,.\label{sec3-def-A-b}
\end{equation}
\end{subequations}
We then obtain
\begin{align}
  \frac{\partial}{\partial x_\nu} \int_y A_\Lambda (x-y) \sigma (y)
  &= \int_y \partial_\nu h_\Lambda (x-y) \int_z \RL (x-y-z) h_\Lambda
    (z) \sigma (y) \nt\\
 &= \int_z \RL (x-z) \frac{\partial}{\partial x_\nu} \int_y h_\Lambda
   (x-y) \sigma (y) h_\Lambda (y-z)\nt\\
   &= \left(\textrm{$\sigma$ term}\right)\,.
\end{align}
Thus, the contribution to $\frac{1}{N} \Theta'_{\mu\nu} (x)$ is
\begin{equation}
\frac{1}{N} \Theta_{\mu\nu}^{\prime (1)} (x) \equiv \delta_{\mu\nu} \int_y A_\Lambda
  (x-y) \sigma (y)\,.
\end{equation}

\subsection{$\sigma^2$ term}

We find
\begin{align}
  \left(\textrm{$\sigma^2$ term of (\ref{sec3-expansion})}\right) 
  &= \int_{z_1, z_2} \sigma (z_1) \sigma (z_2) \int_y \RL (x-y)
    \frac{\partial}{\partial x_\nu} \left( h_\Lambda (x-z_1) h_\Lambda
    (z_1-z_2) h_\Lambda (z_2-y)\right)\nt\\
  &\quad - \frac{1}{2} \sigma (x) \frac{\partial}{\partial x_\nu}
    \int_z h_\Lambda (x-z) \sigma (z) h_\Lambda (z-x) \,.
\end{align}
Using
\begin{equation}
\int_y \RL (x-y) h_\Lambda (y) = \delta (x) + \partial^2 h_\Lambda
(x)\,,
\end{equation}
we obtain
\begin{align}
    (\textrm{$\sigma^2$ term})
  &= \int_{z_1,z_2} \sigma (z_1) \sigma (z_2) \int_y
    \frac{\partial}{\partial x_\nu}
    h_\Lambda (x-z_1) \cdot h_\Lambda (z_1-z_2) \RL (x-y) h_\Lambda (z_2-y)\nt\\
&\quad - \frac{1}{2} \sigma (x) \frac{\partial}{\partial x_\nu} \int_z
  h_\Lambda (x-z) \sigma (z) h_\Lambda (z-x) \nt\\
  &= \int_{z_1, z_2} \sigma (z_1) \sigma (z_2) \partial_\nu h_\Lambda
    (x-z_1) \cdot h_\Lambda (z_1-z_2) \left( \delta (x-z_2) +
    \partial^2 h_\Lambda (x-z_2) \right)\nt\\
  &\quad - \sigma (x) \int_z  \partial_\nu h_\Lambda
    (x-z) \cdot h_\Lambda (z-x) \sigma (z)\nt\\
  &= \frac{1}{2} \int_{z_1, z_2} \sigma (z_1) \sigma (z_2) h_\Lambda
    (z_1-z_2) \nt\\
  &\quad \times \lb \partial_\nu h_\Lambda (x-z_1) \cdot \partial^2
    h_\Lambda (x-z_2) + \partial^2 h_\Lambda (x-z_1) \cdot
    \partial_\nu h_\Lambda (x-z_2) \rb\,.
\end{align}
Thus, the contribution to $\frac{1}{N} \Theta'_{\mu\nu} (x)$ is
\begin{align}
\frac{1}{N} \Theta_{\mu\nu}^{\prime (2)} (x)
  &\equiv  \frac{1}{2} \int_{z_1, z_2} \sigma
    (z_1)  h_\Lambda (z_1-z_2) \sigma (z_2) \nt\\ 
  &\quad \times  \lb - \delta_{\mu\nu} \partial_\alpha h_\Lambda
    (x-z_1) \cdot \partial_\alpha h_\Lambda (x-z_2)\right.\nt\\
  &\quad\left. + \partial_\mu h_\Lambda (x-z_1) \cdot \partial_\nu
    h_\Lambda (x-z_2) + \partial_\nu h_\Lambda (x-z_1) \cdot
    \partial_\mu h_\Lambda (x-z_2) \rb\,.
\end{align}
To see its UV finiteness, it is easier to consider this in the momentum
space:
\begin{align}
  &\int_x e^{- i p x} \frac{1}{N} \Theta_{\mu\nu}^{\prime (2)} (x)
  = \int_{p_1, p_2} \tilde{\sigma} (p_1) \tilde{\sigma}
    (p_2)\, \delta (p_1+p_2 - p)\nt\\
  &\quad\times \frac{1}{2} \int_q \lb
    -\delta_{\mu\nu} q(q+p) + q_\mu (q+p)_\nu + q_\nu (q+p)_\mu \rb
    \tilde{h}_\Lambda (q) \tilde{h}_\Lambda  (q+p_1) \tilde{h}_\Lambda (q+p)\,,
\end{align}
where $\tilde{\sigma} (p)$ is the Fourier transform of $\sigma (x)$.
The integral over $q$ is UV finite for $2 < D < 4$ because
$\tilde{h}_\Lambda (q)$ behaves as $1/q^2$ for large $q$.

Going back to the order $\sigma$ contribution, we might think that 
\begin{equation}
\frac{1}{N} \Theta_{\mu\nu}^{\prime (1)} (x) \equiv  \frac{1}{2} \int_z \sigma (z)
  \lb - \delta_{\mu\nu} \partial_\alpha h_\Lambda (x-z)
  \partial_\alpha h_\Lambda (x-z) + 2 \partial_\mu h_\Lambda (x-z)
  \partial_\nu h_\Lambda (x-z) \rb
\end{equation}
would do.  The integral over $z$ is UV divergent, however.  In
the momentum space we find
\begin{equation}
  \int_x e^{- i p x} \frac{1}{N}
  \Theta_{\mu\nu}^{\prime (1)} (x)
  = \tilde{\sigma} (p) \frac{1}{2} \int_q \lb - \delta_{\mu\nu} q (q+p) +
    q_\mu (q+p)_\nu + q_\nu (q+p)_\mu \rb h_\Lambda (q)  h_\Lambda (q+p),
\end{equation}
which is UV divergent for $D>2$.

\subsection{$\sigma^3$ term}

An analogous calculation gives
\begin{align}
  \frac{1}{N} \Theta_{\mu\nu}^{\prime (3)} (x)
  &= \frac{1}{2} \int_{z_1,z_2,z_3} \sigma (z_1) h_\Lambda (z_1-z_2)
    \sigma (z_2) h_\Lambda (z_2 - z_3) \sigma (z_3)\nt\\
   &\qquad \times  \lb - \delta_{\mu\nu} \partial_\alpha h_\Lambda
    (x-z_1) \cdot \partial_\alpha h_\Lambda (x-z_3)\right.\nt\\
  &\quad\left. + \partial_\mu h_\Lambda (x-z_1) \cdot \partial_\nu
    h_\Lambda (x-z_3) + \partial_\nu h_\Lambda (x-z_1) \cdot
    \partial_\mu h_\Lambda (x-z_3) \rb
\end{align}
as the order $\sigma^3$ contribution.

From the above calculations it is easy to guess the higher order
terms.  We obtain, to all orders of $\sigma$,
\begin{align}
  \frac{1}{N} \Theta'_{\mu\nu} (x)
  &= \delta_{\mu\nu} \int_y A_\Lambda (x-y) \sigma (y)\nt\\
  &\quad + \frac{1}{2} \sum_{n=2}^\infty \int_{z_1,\cdots,z_n} \sigma (z_1)
    h_\Lambda (z_1-z_2) \sigma (z_2) \cdots  h_\Lambda
    (z_{n-1}-z_n) \sigma (z_n) \nt\\
  &\quad\qquad \times  \lb - \delta_{\mu\nu} \partial_\alpha h_\Lambda
    (x-z_1) \cdot \partial_\alpha h_\Lambda (x-z_n)\right.\nt\\
  &\qquad\left. + \partial_\mu h_\Lambda (x-z_1) \cdot \partial_\nu
    h_\Lambda (x-z_n) + \partial_\nu h_\Lambda (x-z_1) \cdot
    \partial_\mu h_\Lambda (x-z_n) \rb\nt\\
  &= \delta_{\mu\nu} \int_y A_\Lambda (x-y) \sigma (y)\nt\\
  &\quad + \frac{1}{2} \lim_{y \to x}
    \left( - \delta_{\mu\nu} \frac{\partial}{\partial x_\alpha}
    \frac{\partial}{\partial y_\alpha} + \frac{\partial}{\partial
    x_\mu} \frac{\partial}{\partial y_\nu} + \frac{\partial}{\partial
    x_\nu} \frac{\partial}{\partial y_\mu} \right)\nt\\
  &\qquad \times \lb \G_\Lambda (x,y) [\sigma] - h_\Lambda (x-y) -
    \int_z h_\Lambda (x-z) \sigma (z) h_\Lambda (z-y) \rb\,.
\end{align}
In conclusion we obtain, from (\ref{sec3-EMprime}) and the above,
\begin{align}
  \Theta_{\mu\nu} (x)
  &= \partial_\mu \Phi^I (x) \partial_\nu \Phi^I (x) - \frac{1}{2}
    \delta_{\mu\nu} \partial_\alpha \Phi^I (x) \partial_\alpha \Phi^I
    (x) - N \delta_{\mu\nu} \frac{f_2}{2} \sigma (x)^2\nt\\
  &\quad + N \Bigg[
    \delta_{\mu\nu} \left(\alpha_\Lambda + \int_y A_\Lambda (x-y)
    \sigma (y) \right) + \left(- \delta_{\mu\nu} \partial^2 +
    \partial_\mu \partial_\nu\right) \int_y \beta_\Lambda (x-y) \sigma
    (y)\nt\\
  &\qquad + \frac{1}{2} \lim_{y \to x}
    \left( - \delta_{\mu\nu} \frac{\partial}{\partial x_\alpha}
    \frac{\partial}{\partial y_\alpha} + \frac{\partial}{\partial
    x_\mu} \frac{\partial}{\partial y_\nu} + \frac{\partial}{\partial
    x_\nu} \frac{\partial}{\partial y_\mu} \right)\nt\\
  &\qquad\qquad \times \lb\G_\Lambda (x,y) [\sigma] - h_\Lambda (x-y) -
    \int_z h_\Lambda (x-z) \sigma (z) h_\Lambda (z-y) \rb \Bigg]\,,
    \label{sec3-EM}
\end{align}
where the function $A_\Lambda (x)$ is defined by (\ref{sec3-def-A}).
Please note that we have introduced two extra terms, a constant
proportional to $\delta_{\mu\nu}$ and another proportional to
$\delta_{\mu\nu} \partial^2 - \partial_\mu \partial_\nu$.  Neither
contributes to $\partial_\mu \Theta_{\mu\nu} (x)$, and we cannot
determine their coefficients $\alpha_\Lambda$ and
$\beta_\Lambda (x-y)$ at this point.

In the next section we review and discuss cutoff dependent composite
operators in large $N$.  We will be able to determine both
$\alpha_\Lambda$ and $\beta_\Lambda (x-y)$ by demanding that
$\Theta_{\mu\nu} (x)$ have the correct cutoff dependence.\footnote{As
  we will explain in Sec.~\ref{sec-completion}, $\beta_\Lambda (x-y)$
  is unique up to the addition of a constant multiple of
  $f_2 \delta (x-y)$.}

\section{Cutoff dependent composite operators in large
  $N$\label{sec-comp}}

To introduce cutoff dependent composite operators,\footnote{These were
  first introduced in \cite{Becchi:1996an}.  Here we adopt a
  pedagogical approach.}  we go back to the cutoff dependent
generating functional $W_\Lambda [J]$ defined by
\begin{equation}
  e^{W_\Lambda [J]} = \int [d\phi] \exp \left[ \Sb [\phi] -
    \frac{1}{2} \int_{x,y} \RL (x-y) \phi (x) \phi (y) + \int_x J(x)
    \phi (x) \right]\,.
\end{equation}
We consider a single field $\phi$ since the generalization to $N$
fields is straightforward.  An infinitesimal change of the bare action
$\Sb [\phi]$ by $\Op_\mathrm{bare} [\phi]$ induces an infinitesimal
change $\Op_\Lambda [J]$ of $W_\Lambda [J]$ by
\begin{equation}
  \Op_\Lambda [J] e^{W_\Lambda [J]}
  = \int [d\phi] \,\Op_\mathrm{bare} [\phi]  \exp \left[ \Sb [\phi] -
    \frac{1}{2} \int_{x,y} \RL (x-y) \phi (x) \phi (y) + \int_x J(x)
    \phi (x) \right]\,.\label{sec4-OL-J}
\end{equation}
Regarding $\Op_\Lambda [J]$ as a functional of $\Phi (x) \equiv
\frac{\delta W_\Lambda [J]}{\delta J(x)}$, we obtain a cutoff
dependent composite operator $\Op_\Lambda [\Phi]$ (using the same
symbol).  The cutoff dependence of $\Op_\Lambda [\Phi]$ is obtained as
\begin{equation}
  - \Lambda \frac{\partial}{\partial \Lambda} \Op_\Lambda [\Phi]
  = \frac{1}{2} \int_{x,y} \RL (x-y) \int_{x',y'} \G_\Lambda (x,x') [\Phi] \frac{\delta}{\delta
    \Phi (x')} \Op_\Lambda [\Phi] \frac{\overleftarrow{\delta}}{\delta
    \Phi (y')} \G_\Lambda (y,y') [\Phi]\,,\label{sec4-ERG-OL}
\end{equation}
where $\G_\Lambda (x,y) [\Phi]$ is defined by
(\ref{sec2-GL}).\footnote{We derive this in Appendix \ref{appendix-A}.}  The simplest
example is $\Op_\Lambda [\Phi] = \Phi (x)$, which corresponds to
$\Op_\mathrm{bare} [\phi] = \phi (x)$.  The EM tensor
$\Theta_{\mu\nu} (x)$ is another example of a cutoff dependent
composite operator for the choice
$\Op_\mathrm{bare} [\phi] = \Theta^{\mathrm{bare}}_{\mu\nu} (x)$.

Note that $\Op_\Lambda [\Phi]$ is a functional of $\Phi$.  In the
limit $\Lambda \to 0+$, we obtain the 1PI correlation functions:
\begin{equation}
  \lim_{\Lambda \to 0+} \Op_\Lambda [\Phi]
  = \sum_{n=0}^\infty \frac{1}{n!} \int_{x_1, \cdots, x_n} \Phi (x_1)
  \cdots \Phi (x_n)\, \vev{\Op\, \phi (x_1) \cdots \phi
    (x_n)}^{\mathrm{1PI}}\,,
\end{equation}
where the suffix 1PI denotes the 1PI part of the correlation function
\begin{equation}
  \vev{\Op\, \phi (x_1) \cdots \phi (x_n)}
  \equiv \int [d\phi] \, \Op_\mathrm{bare} [\phi] \phi (x_1) \cdots
  \phi (x_n) \, e^{\Sb[\phi]}\,.
\end{equation}

In large $N$  we find
\begin{equation}
  \G^{IJ}_{\Lambda} (x,y) [\Phi] \simeq \delta^{IJ} \G_{\Lambda} (x,y)
  [\sigma]
\end{equation}
which, regarded as a functional of $\sigma$, is determined by
(\ref{sec2-GL-largeN}).  Hence, in large $N$, the cutoff dependence
of a composite operator is given by
\begin{equation}
  - \Lambda \partial_\Lambda \Op_\Lambda [\Phi]
  = \frac{1}{2} \int_{x,y} \Lambda \partial_\Lambda \RL (x-y)
  \int_{x',y'}  \G_{\Lambda} (x,x') [\sigma] \G_{\Lambda} (y,y') [\sigma]
  \frac{\delta^2 \Op_\Lambda [\Phi]}{\delta \Phi^I (x') \delta 
    \Phi^I (y')}\, .\label{sec4-ERG-comp}
\end{equation}

If $\Op_\Lambda [\Phi]$ depends only on $\rho$, this ERG equation
can be simplified further.  Since
\begin{equation}
  \frac{\delta \Op_\Lambda [\rho]}{\delta \Phi^J (x)} 
= \frac{\delta \Op_\Lambda [\rho]}{\delta \rho (x)} \frac{1}{N} \Phi^J (x)\,,
\end{equation}
we obtain
\begin{align}
  \frac{\delta^2 \Op_\Lambda [\rho]}{\delta \Phi^J (x) \delta \Phi^K (y)} 
  &= \frac{1}{N} \delta^{JK} \delta (x-y) \frac{\delta \Op_\Lambda
    [\rho]}{\delta \rho (x)} 
 + \frac{1}{N^2}  \Phi^J (x) \Phi^K (y)
    \frac{\delta^2 \Op_\Lambda [\rho]}{\delta \rho (x) \delta \rho (y)}\nt\\
&\simeq \frac{1}{N} \delta^{JK} \delta (x-y) \frac{\delta \Op_\Lambda
    [\rho]}{\delta \rho (x)}\,.
\end{align}
Hence, in large $N$, the ERG equation is simplified as
\begin{equation}
 - \Lambda \frac{\partial}{\partial \Lambda} \Op_\Lambda [\rho]
  = \frac{1}{2} \int_{x,y} \Lambda \partial_\Lambda \RL (x-y)
  \int_z \G_\Lambda (x,z) [\sigma]\frac{\delta \Op_\Lambda [\rho]}{\delta \rho
    (z)} \G_\Lambda (z,y) [\sigma]\,.\label{sec4-ERG-comp-largeN}
\end{equation}
Note that the right-hand side involves only the first order
differential with respect to $\rho$.
Composite operators in the $O(N)$ model have been studied before in Refs.~\cite{Rose:2015bma,Rose:2016elj,Rose:2021zdk,Cabrera:2024rgy}.

\subsection{$\sigma$ as a cutoff dependent composite operator}

Let us show that $\sigma (x)$ and its products are cutoff dependent
composite operators satisfying (\ref{sec4-ERG-comp-largeN}).  By
definition (\ref{sec2-sigma}), $\sigma (x)$ is a functional of $\rho$.
In the large $N$ approximation (\ref{sec2-ERG-full}) reduces to
\begin{equation}
  - \Lambda \partial_\Lambda \Gamma_{I\Lambda} [\rho]
  = \frac{1}{2} \int_{x,y} \Lambda \partial_\Lambda \RL (x-y) 
  \G_{\Lambda} (x,y) [\sigma]\,.
\end{equation}
Differentiating this with respect to $\rho (x)$, we obtain
\begin{equation}
  - \Lambda \partial_\Lambda \sigma (x)\Big|_{\rho\,\mathrm{fixed}}
  = \frac{1}{2} \int_{y,z} \Lambda \partial_\Lambda \RL (y-z) \cdot
  \frac{\delta}{\delta \rho (x)} \G_{\Lambda} (y,z) [\sigma]\,,
\end{equation}
where $\sigma$ is regarded as a functional of $\rho$ on the left-hand
side.  Differentiating (\ref{sec2-GL-largeN}) with respect to
$\rho (x)$, we obtain
\begin{equation}
  \int_z \frac{\delta \G_{\Lambda} (y,z) [\sigma]}{\delta \rho (x)} \lb
  - \left(\partial_z^2 + \sigma (z) \right) \delta (z-w) + \RL (z-w) \rb
  - \G_\Lambda (y,w) \frac{\delta^2 \Gamma_{I\Lambda}}{\delta \rho (w)
    \delta \rho (x)}  = 0\,.
\end{equation}
Using (\ref{sec2-GL-largeN}), we can solve this as
\begin{equation}
  \frac{\delta \G_{\Lambda} (y,z) [\sigma]}{\delta \rho (x)}
  = \int_{w} \G_{\Lambda} (y,w) [\sigma] \frac{\delta^2 \Gamma_{I\Lambda}
    [\rho]}{\delta \rho (w) \delta \rho (x)} \G_{\Lambda} (w,z) [\sigma]\,.
\end{equation}
Hence, we obtain
\begin{equation}
  - \Lambda \partial_\Lambda \sigma (x)\Big|_{\rho\,\textrm{fixed}}
  = \frac{1}{2} \int_{y,z} \Lambda \partial_\Lambda \RL (y-z)
  \int_{w} \G_{\Lambda} (y,w) [\sigma] \frac{\delta \sigma
    (x)}{\delta \rho (w)} \G_{\Lambda} (w,z) [\sigma]\,.\label{sec4-ERG-sigma}
\end{equation}
Thus, as a functional of $\rho$, $\sigma (x)$ is a cutoff dependent
composite operator.  Moreover, any product of $\sigma$'s
\begin{equation}
  \Op_\Lambda [\sigma] \equiv  \int_{x_1, \cdots, x_n}  C_n (x_1, \cdots,
  x_n)\, \sigma (x_1) \cdots  \sigma (x_n)
\end{equation}
is also a composite operator as long as the coefficient $C_n$ has no
$\Lambda$ dependence, because (\ref{sec4-ERG-comp-largeN}) involves
only the first order differential with respect to $\rho$.  For
example, both
\begin{equation}
  \frac{\partial}{\partial f_1} F_\Lambda [\sigma] = \int_x \sigma (x)
\end{equation}
and
\begin{equation}
  \frac{\partial}{\partial f_2} F_\Lambda [\sigma] = \frac{1}{2}
  \int_x \sigma (x)^2
\end{equation}
are cutoff dependent composite operators.

\subsection{A class of cutoff dependent composite
  operators\label{sec4-subsec-class}}

Let 
\begin{equation}
\Op_\Lambda (x) [\Phi] \equiv \frac{1}{2} \int_{x_1, x_2} C (x-x_1, x-x_2)
\left( \frac{1}{N} \Phi^I (x_1) \Phi^I (x_2) + \G_\Lambda (x_1, x_2) [\sigma]
\right)\,.
\label{sec4-comp-quad}
\end{equation}
As the coefficient,  we take
\begin{equation}
  C(x-x_1, x-x_2) = \int_{p_1, p_2} e^{i p_1 (x-x_1) + i p_2 (x-x_2)}
  \tilde{C} (p_1, p_2)\,,
\end{equation}
where $\tilde{C}$ is a symmetric polynomial of $p_1, p_2$ independent
of $\Lambda$.  In other words $C$ is a sum of derivatives of
$\delta (x-x_1) \delta (x-x_2)$ with respect to $x_1, x_2$.

We wish to show that (\ref{sec4-comp-quad}) is a $\Lambda$-dependent
composite operator satisfying (\ref{sec4-ERG-comp}).  As a
preparation, we compute
\begin{align}
  - \Lambda \partial_\Lambda \G_{\Lambda} (x,y)  [\sigma]\Big|_{\rho\,\textrm{fixed}} 
  &= - \Lambda \partial_\Lambda \G_{\Lambda} (x,y)  [\sigma]\Big|_{\sigma\,\textrm{fixed}}
  + \int_z \left(- \Lambda \partial_\Lambda \sigma
    (z)\right)_{\rho\,\textrm{fixed}} \frac{\delta \G_{\Lambda}
    (x,y)[\sigma]}{\delta \sigma (z)}\nt\\
  &= - \Lambda \partial_\Lambda \G_{\Lambda; -p,q} (x,y)
    [\sigma]\Big|_{\sigma\,\textrm{fixed}}  \nt\\
  &\quad + \frac{1}{2}
\int_{u,v} \Lambda \partial_\Lambda \RL (u-v) \int_z \G_\Lambda (u,z)
    \frac{\delta \G_\Lambda (x,y)}{\delta \rho (z)} \G_\Lambda (z,v)\,,
\end{align}
where we have used (\ref{sec4-ERG-sigma}).
Differentiating
(\ref{sec2-GL-largeN}) with respect to $\Lambda$, we obtain
\begin{align}
&   \int_y (- \Lambda \partial_\Lambda \G_{\Lambda} (x,y)
    [\sigma]) \lb \left(- \partial_y^2 - \sigma (y)\right) \delta
    (y-z) + \RL (y-z) \rb\nt\\
&\quad = \int_y \G_{\Lambda} (x,y) [\sigma]  \Lambda \partial_\Lambda \RL (y-z)\,.
\end{align}
This gives
\begin{equation}
 - \Lambda \partial_\Lambda \G_{\Lambda} (x,y)
 [\sigma]\Big|_{\sigma\,\textrm{fixed}}
 = \int_{u,v} \, \G_\Lambda (x,u) \Lambda \partial_\Lambda \RL (u-v)
 \G_\Lambda (v,y)\,.
\end{equation}
Hence,
\begin{align}
  - \Lambda \partial_\Lambda \G_{\Lambda} (x,y) [\sigma] \Big|_{\rho\,\textrm{fixed}}
    &= \int_{u,v} \G_{\Lambda} (x,u) [\sigma] \Lambda \partial_\Lambda
      \RL (u-v)  \G_{\Lambda} (v,y) [\sigma]\nt\\
    &\quad + \frac{1}{2} \int_{u,v} \Lambda \partial_\Lambda \RL (u-v)
      \int_z \G_\Lambda (u,z) \frac{\delta \G_\Lambda (x,y)}{\delta
      \rho (z)} \G_\Lambda (z,v)\,.
\end{align}
We therefore obtain
\begin{align}
  - \Lambda \partial_\Lambda \Op_\Lambda (x) [\Phi]
  &= \frac{1}{2} \int_{x_1, x_2}  C (x-x_1, x-x_2) 
    (-) \Lambda \partial_\Lambda \G_{\Lambda} (x_1, x_2) \Big|_{\rho\,\textrm{fixed}}\nt\\
  &= \frac{1}{2} \int_{x_1, x_2}  C (x-x_1, x-x_2) 
    \int_{u,v} \Lambda \partial_\Lambda \RL (u-v) \nt\\
  &\quad \times \lb \G_{\Lambda} (x_1, u) \G_\Lambda (v,x_2)
 +  \frac{1}{2}
    \int_z \G_\Lambda (u,z) \frac{\delta \G_\Lambda (x_1, x_2)}{\delta
    \rho (z)} \G_\Lambda (z,v) \rb\,.
\end{align}
We now compare this with
\begin{align}
  &\frac{1}{2} \int_{u,v} \Lambda \partial_\Lambda \RL (u-v)
\int_{z,w} \G_\Lambda (u,z) \G_\Lambda (v,w) \frac{\delta^2
    \Op_\Lambda (x)}{\delta \Phi^I (z) \delta \Phi^I (w)}\nt\\
  &\simeq  \frac{1}{2} \int_{u,v} \Lambda \partial_\Lambda \RL (u-v)
    \int_{z,w} \G_\Lambda (u,z) \G_\Lambda (v,w)\nt\\
  &\quad\times \int_{x_1, x_2} C(x-x_1, x-x_2)
    \left( \delta (z-x_1) \delta (w-x_2) +
    \frac{1}{2}\frac{\delta^2}{\delta \Phi^I (z) \delta \Phi^I (w)}
    \G_\Lambda (x_1, x_2) [\sigma]\right)\nt\\
  &\simeq \frac{1}{2} \int_{u,v} \Lambda \partial_\Lambda \RL (u-v)
    \int_{x_1, x_2} C (x-x_1, x-x_2)\nt\\
  &\quad \times \lb \G_\Lambda (u, x_1) \G_\Lambda (v,x_2) +
    \frac{1}{2} \int_z \G_\Lambda (u,z) \G_\Lambda (v,z) \frac{\delta
    \G_\Lambda (x_1, x_2)}{\delta \rho (z)} \rb\,,
\end{align}
where we have used
\begin{equation}
  \frac{\delta^2}{\delta \Phi^I (z) \delta \Phi^I (w)} \G_\Lambda
  (x_1, x_2) [\sigma]
  \simeq \delta (z-w) \frac{\delta \G_\Lambda (x_1, x_2)}{\delta \rho
    (w)}\,,
\end{equation}
valid in large $N$.  Hence, $\Op_\Lambda (x) [\Phi]$ has the correct
$\Lambda$ dependence (\ref{sec4-ERG-comp}).
  
There is one subtlety, however.  The integral
\begin{equation}
  \frac{1}{2} \int_{x_1, x_2} C (x-x_1, x-x_2) \G_\Lambda (x_1, x_2) [\sigma]
\end{equation}
may not converge if $C$ involves too many derivatives.  Suppose
$C(x-x_1, x-x_2)$ involves two derivatives so that
$\tilde{C} (p_1, p_2)$ is a quadratic polynomial.  Expanding in powers
of $\sigma$, we obtain
\begin{align}
  & \int_x e^{- i p x} \frac{1}{2} \int_{x_1, x_2} C (x-x_1, x-x_2)
    \G_\Lambda (x_1, x_2) [\sigma] \nt\\
  &=  \frac{1}{2} \int_{p_1, p_2} \delta (p_1+p_2-p) \tilde{C} (p_1, p_2) \Bigg[
    \tilde{h}_\Lambda (p_1) \delta (p) \nt\\
  &\quad + \tilde{h}_\Lambda (p_1) \lb \tilde{\sigma} (p)
   + \int_{q_1, q_2} \delta (q_1+q_2-p) \tilde{\sigma} (q_1) \tilde{h}_\Lambda
    (-p_1+q_1) \tilde{\sigma} (q_2) + \cdots \rb \tilde{h}_\Lambda (p_2) \Bigg].
\end{align}
The first two terms
\begin{subequations}
\begin{align}
  a_\Lambda 
  &= \frac{1}{2} \int_{p_1, p_2} \delta (p_1+p_2) \tilde{C} (p_1, p_2) \tilde{h}_\Lambda (p_1)
  = \frac{1}{2} \int_q \tilde{C} (-q,q) \tilde{h}_\Lambda (q),\\
  \tilde{b}_\Lambda (p)
  &= \frac{1}{2} \int_{p_1, p_2} \delta (p_1+p_2-p) \tilde{C} (p_1,
    p_2) \tilde{h}_\Lambda (p_1)  \tilde{h}_\Lambda (p_2)
    = \frac{1}{2} \int_q \tilde{C} (-q,q+p) \tilde{h}_\Lambda (q) \tilde{h}_\Lambda (q+p) 
\end{align}
\end{subequations}
may be UV divergent for $2 < D < 4$.  But we can still define them as
the solution to the differential equations
\begin{subequations}\label{sec4-diffeq-ab}
\begin{align}
  - \Lambda \partial_\Lambda a_\Lambda
  &= \frac{1}{2} \int_q \tilde{C}(-q,q) \tilde{f}_\Lambda (q), \\
  - \Lambda \partial_\Lambda \tilde{b}_\Lambda (p)
  &= \frac{1}{2} \int_q \tilde{C} (-q, q+p)\,   \left(
    \tilde{f}_\Lambda (q) \tilde{h}_\Lambda (q+p) + 
    \tilde{h}_\Lambda (q) \tilde{f}_\Lambda (q+p) \right) ,
\end{align}
\end{subequations}
where
\begin{equation}
  \tilde{f}_\Lambda (q) \equiv - \Lambda \partial_\Lambda \tilde{h}_\Lambda (q) =
  \Lambda \partial_\Lambda R_\Lambda (q) \cdot \tilde{h}_\Lambda (q)^2 .
\end{equation}
The right-hand sides of (\ref{sec4-diffeq-ab}) are absolutely
convergent because $\tilde{f}_\Lambda (q)$ decays fast for large $q$.
  
\subsection{An example}

We consider the simplest example from the class of composite operators
discussed in Sec.~B above.  We choose
\begin{equation}
  C (x_1, x_2) = \delta (x-x_1) \delta (x-x_2)
\end{equation}
to obtain
\begin{align}
\left[\rho (x)\right]_\Lambda
  &\equiv \frac{1}{2N} \Phi^I (x) \Phi^I (x) + a_\Lambda + \frac{1}{2}
    \int_z h_\Lambda (x-z) \sigma (z) h_\Lambda (z-x)\nt\\
&\quad +\frac{1}{2} \lim_{x' \to x} \lb \G_\Lambda (x,x') -
    h_\Lambda (x-x')  - \int_z h_\Lambda (x-z) \sigma (z) h_\Lambda
    (z-x') \rb\nt\\
  &= \frac{1}{2N} \Phi^I (x) \Phi^I (x) + a_\Lambda 
 +\frac{1}{2} \lim_{x' \to x} \lb \G_\Lambda (x,x') -
    h_\Lambda (x-x') \rb\,,\label{sec4-comp-rho}
\end{align}
where $a_\Lambda$ is determined by
\begin{equation}
  - \Lambda \partial_\Lambda a_\Lambda
  = \frac{1}{2} \int_q \tilde{f}_\Lambda (q)\,,\label{sec4-diffeq-a}
\end{equation}
and
\begin{equation}
  \int_z h_\Lambda (x-z) \sigma (z) h_\Lambda (z-x)
\end{equation}
is finite.  Solving (\ref{sec4-diffeq-a}) we obtain
\begin{equation}
  a_\Lambda =  \frac{1}{2-D} \frac{1}{2}
  \int_q \tilde{f}_\Lambda (q) = \frac{1}{2} \int_q \left( \tilde{h}_\Lambda (q) -
    \frac{1}{q^2}\right) = c_{1\Lambda},\label{sec4-rho-a}
\end{equation}
where $c_{1\Lambda}$ is defined in (\ref{sec2-c1Lambda}).

We now show that $[\rho (x)]_\Lambda$ can in fact be written in terms
of $\sigma$ only, and that it vanishes at $f_1=f_2=0$.  We recall that
the inverse Legendre transformation from $F_\Lambda [\sigma]$ to
$\Gamma_{I\Lambda} [\rho]$ gives
\begin{equation}
  \rho (x) = - \frac{\delta F_\Lambda [\sigma]}{\delta \sigma (x)}
  = - f_1  - f_2 \sigma (x) - \frac{\delta I_\Lambda
    [\sigma]}{\delta \sigma (x)}\,,
\end{equation}
where $I_\Lambda [\sigma]$ is defined by (\ref{sec2-def-ILambda}).
Hence, we obtain
\begin{align}
&  - f_1  - f_2 \sigma (x)
  = \rho (x) + \frac{\delta I_\Lambda
    [\sigma]}{\delta \sigma (x)}\nt\\
  &\quad= \rho (x) + c_{1\Lambda}  + \frac{1}{2}
    \sum_{n=1}^\infty
    \int_{x_1,\cdots,x_n} \sigma (x_1) \cdots \sigma
    (x_n)\,I_{n\Lambda} (x, x_1, \cdots, x_n)\nt\\
  &\quad= \rho (x) + c_{1\Lambda}  + \frac{1}{2}
    \lim_{x'\to x} \lb \G_{\Lambda} (x,x') - h_\Lambda (x-x') \rb\nt\\
  &\quad= \left[\rho (x)\right]_\Lambda \,.\label{sec4-rho-sigma}
\end{align}
Generalizing this result, we find that any $\Lambda$-dependent
composite operator written in terms of $\rho$ can be expressed in
terms of $\sigma$.  Those involving a term like
$\partial_\mu \Phi^I \partial_\nu \Phi^I$ are not of this type.

\subsection{Equation of motion operators}

There is a special class of cutoff dependent composite operators,
called equation of motion operators.\cite{Becchi:1996an,
  Igarashi:2009tj,Dietz:2013sba} They correspond to bare fields of the
type
\begin{equation}
  \mathcal{E}_\mathrm{bare} (x) = - e^{- \Sb [\phi]} \frac{\delta}{\delta \phi
    (x)} \left( \Op_\mathrm{bare} [\phi] e^{\Sb [\phi]} \right)\,.
\end{equation}
This has the correlation function
\begin{align}
  \vev{\mathcal{E} (x)\, \phi (x_1) \cdots \phi (x_n)}
&\equiv \int [d\phi] \, \mathcal{E}_\mathrm{bare} (x) \, \phi (x_1) \cdots
  \phi (x_n)\,e^{\Sb [\phi]}\nt\\
  &= \int [d\phi] \,\Op_\mathrm{bare} [\phi] \frac{\delta}{\delta \phi
    (x)} \lb \phi (x_1) \cdots \phi (x_n)\rb \,e^{\Sb [\phi]}\nt\\
  &= \sum_{i=1}^n \delta (x-x_i) \vev{\Op\, \phi (x_1) \cdots
    \widehat{\phi (x_i)} \cdots \phi (x_n)}\,,\label{sec4-eom-correlation}
\end{align}
where $\phi (x_i)$ is replaced by $\Op$.  These operators were
originally introduced by Wegner \cite{Wegner:1974sla} and called
redundant operators since they only introduce change of field
variables but keep the theory unchanged.

Corresponding to $\mathcal{E}_\mathrm{bare} (x)$, we obtain the cutoff
dependent composite operator $\mathcal{E}_\Lambda (x)$:
\begin{align}
  \mathcal{E}_\Lambda (x) \,e^{W_\Lambda [J]}
  &\equiv \int [d\phi]\, \mathcal{E}_\mathrm{bare} (x) \exp \left[ \Sb
    [\phi] - \frac{1}{2} \int_{x,y} \RL (x-y) \phi (x) \phi (y) +
    \int_x J(x) \phi (x) \right]\nt\\
  &= \int [d\phi]\, \Op_\mathrm{bare} [\phi] \left( - \int_y \RL (x-y) \phi (y) +
    J(x) \right) \nt\\
  &\quad \times \exp \left[ \Sb
    [\phi] - \frac{1}{2} \int_{x,y} \RL (x-y) \phi (x) \phi (y) +
    \int_x J(x) \phi (x) \right]\nt\\
  &= \lb  J(x) \Op_\Lambda [J] - \int_y \RL (x-y)
    \left( \frac{\delta \Op_\Lambda [J]}{\delta J(y)} + \Op_\Lambda
    [J] \frac{\delta W_\Lambda [J]}{\delta J(y)} \right) \rb e^{W_\Lambda
    [J]} \,,
\end{align}
where $\Op_\Lambda$ is the cutoff dependent composite operator given
by (\ref{sec4-OL-J}).  Regarding $\mathcal{E}_\Lambda (x)$ as a
functional of
$\Phi (x) \equiv \frac{\delta W_\Lambda [J]}{\delta J(x)}$, we obtain
\begin{equation}
  \mathcal{E}_\Lambda (x) [\Phi]
  = - \Op_\Lambda [\Phi] \frac{\delta \Gamma_\Lambda [\Phi]}{\delta \Phi
    (x)} - \int_{y,z} \RL (x-y) \G_\Lambda (y,z) [\Phi] \frac{\delta
    \Op_\Lambda [\Phi]}{\delta \Phi (z)}\,.
\end{equation}
The choice $\Op_\mathrm{bare} [\phi] = \phi (x)$ gives the simplest
example of $\mathcal{E}_\Lambda [\Phi]$:
\begin{equation}
  \N_\Lambda (x) \equiv - \Phi (x) \frac{\delta \Gamma_\Lambda
    [\Phi]}{\delta \Phi (x)} - \int_y \RL (x-y) \G_\Lambda (x,y) [\Phi]\,.
\end{equation}
We call this a number operator because it counts the number of scalar
fields in the correlation function (\ref{sec4-eom-correlation}).

For the linear sigma model, the number operator is generalized to
\begin{equation}
  \N_\Lambda (x) \equiv - \Phi^I (x) \frac{\delta \Gamma_\Lambda
    [\Phi]}{\delta \Phi^I (x)} - \int_y \RL (x-y)\,\G_\Lambda^{II}
  (x,y) [\Phi]\,.
\end{equation}
In large $N$, we obtain
\begin{equation}
  \N_\Lambda (x) = - \Phi^I (x) \partial^2 \Phi^I (x) - 2 N \rho (x)
  \sigma (x) - N \int_y \RL (x-y) \G_\Lambda (x,y) [\sigma]\,.
\end{equation}
By construction this should have the correct cutoff dependence given
by (\ref{sec4-ERG-comp}).  In the remaining part we show that
$\N_\Lambda (x)$ consists of terms proportional to
$\sigma (x), \sigma (x)^2$, and one of the cutoff dependent composite
operators discussed in Sec.~\ref{sec4-subsec-class}.

Using
\begin{equation}
  \rho (x) = - \frac{\delta F_\Lambda [\sigma]}{\delta \sigma (x)}
  = - f_1 - f_2 \sigma (x) - \frac{\delta I_\Lambda [\sigma]}{\delta
    \sigma (x)}\,,
\end{equation}
we obtain
\begin{align}
  \N_\Lambda (x)
  &= - \Phi^I (x)\partial^2 \Phi^I (x) + 2 N \left( f_1 \sigma (x) +
    f_2 \sigma (x)^2 \right) \nt\\
  &\quad + 2 N \sigma (x) \frac{\delta
    I_\Lambda}{\delta \sigma (x)} - N \int_y \RL (x-y) \G_\Lambda
    (x,y) [\sigma]\,.
\end{align}
Since
\begin{align}
  & - 2 \sigma (x)
    \frac{\delta I_\Lambda [\sigma]}{\delta \sigma (x)}
    + \int_y \RL (x-y) \G_\Lambda (x,y) [\sigma] \nt\\
  &= \int_y \RL (y) h_\Lambda (y) + \int_{y,z} \RL (x-y) h_\Lambda
    (x-z) \sigma (z) h_\Lambda (z-y) - 2 c_{1\Lambda} \sigma (x)\nt\\
  &\quad + \lim_{y \to x} \frac{1}{2} \left(\partial_x^2 +
    \partial_y^2\right)
    \left( \G_\Lambda (x,y) [\sigma] - h_\Lambda (x-y) - \int_z
    h_\Lambda (x-z) \sigma (z) h_\Lambda (z-y) \right)\,,
\end{align}
where (\ref{sec2-diffeq-hL}) is used, we obtain
\begin{align}\label{sec4-number-expansion}
  &\N_\Lambda (x)
  = 2 N \left( f_1 \sigma (x) + f_2 \sigma (x)^2 \right)\nt\\
  &\quad - \Phi^I (x) \partial^2 \Phi^I (x) 
 + N  \left[ a_\Lambda + \int_y b_\Lambda (x-y) \sigma (y)\right.\\
&\left.\qquad  - \lim_{y\to x} \frac{1}{2} \left(\partial_x^2 +
    \partial_y^2\right)
    \left( \G_\Lambda (x,y) [\sigma] - h_\Lambda (x-y) - \int_z
  h_\Lambda (x-z) \sigma (z) h_\Lambda (z-y) \right)\right]\,,\nt
\end{align}
where
\begin{subequations}\label{sec4-number-ab}
  \begin{align}
  a_\Lambda
  &\equiv - \int_y \RL (y) h_\Lambda (y) = - \int_q R_\Lambda (q)
    \tilde{h}_\Lambda (q)\,,\label{sec4-number-a}\\
  b_\Lambda (x-y)
  &\equiv 2 c_{1\Lambda} \delta (x-y) - h_\Lambda (x-y) \int_z \RL
  (x-z) h_\Lambda (z-y)\,.
  \end{align}
\end{subequations}
The part starting with $- \Phi^I \partial^2 \Phi^I$ must belong to the
class of composite operators in Sec.~\ref{sec4-subsec-class},
corresponding to $\tilde{C} (p_1, p_2) = p_1^2 + p_2^2$.  Hence,
$a_\Lambda$ and
\begin{equation}
  \tilde{b}_\Lambda (p) \equiv
  \int_x e^{- i p x} b_\Lambda (x) = 2 c_{1\Lambda} - \int_q  R_\Lambda (q)
  \tilde{h}_\Lambda (q) \tilde{h}_\Lambda (q+p) 
\label{sec4-number-b}
\end{equation}
must satisfy
\begin{subequations}
  \begin{align}
    - \Lambda \partial_\Lambda a_\Lambda
    &= \int_q q^2 \tilde{f}_\Lambda (q),\label{sec4-number-a-diff}\\
    - \Lambda \partial_\Lambda \tilde{b}_\Lambda (p)
    &= \int_q q^2 \left( \tilde{f}_\Lambda (q) \tilde{h}_\Lambda (q+p) + \tilde{f}_\Lambda
      (q+p) \tilde{h}_\Lambda (q)\right),\label{sec4-number-b-diff}
  \end{align}
\end{subequations}
respectively.

Let us check (\ref{sec4-number-a-diff}) first.  Differentiating
(\ref{sec4-number-a}), we find
\begin{align}
  - \Lambda \partial_\Lambda a_\Lambda
  &= \int_q \left( \Lambda \partial_\Lambda R_\Lambda (q) \cdot
    \tilde{h}_\Lambda (q) - R_\Lambda (q) \tilde{f}_\Lambda (q) \right)\nt\\
  &= \int_q \tilde{f}_\Lambda (q) \left( \frac{1}{\tilde{h}_\Lambda (q)} - R_\Lambda
    (q) \right) = \int_q q^2 \tilde{f}_\Lambda (q),
\end{align}
verifying (\ref{sec4-number-a-diff}).  We next
differentiate (\ref{sec4-number-b}) to find
\begin{align}
  - \Lambda \partial_\Lambda \tilde{b}_\Lambda (p)
  &= \int_q \left[ \Lambda \partial_\Lambda R_\Lambda (q) \cdot
    \tilde{h}_\Lambda (q) \tilde{h}_\Lambda (q+p) \right.\nt\\
  &\left.\quad - R_\Lambda (q) \left(\tilde{f}_\Lambda (q)
    \tilde{h}_\Lambda (q+p) + \tilde{h}_\Lambda (q) \tilde{f}_\Lambda (q+p) \right)
 +  \tilde{f}_\Lambda (q)\right]\nt\\
  &= \int_q \left[ \tilde{f}_\Lambda (q) q^2 \tilde{h}_\Lambda
    (q+p) + \tilde{f}_\Lambda (q+p) 
    \left(-1 + q^2 \tilde{h}_\Lambda (q)\right) +
    \tilde{f}_\Lambda (q+p) \right]\nt\\ 
  &= \int_q q^2 \left( \tilde{f}_\Lambda (q) \tilde{h}_\Lambda
    (q+p) + \tilde{h}_\Lambda (q)   \tilde{f}_\Lambda (q+p) \right),
\end{align}
thus verifying (\ref{sec4-number-b-diff}).

\section{Completion of the energy-momentum
  tensor\label{sec-completion}}

In Sec.~\ref{sec-EM} we have constructed the energy-momentum tensor
$\Theta_{\mu\nu} (x)$ as given by (\ref{sec3-EM}).  We are still
missing two coefficients $\alpha_\Lambda$ and $\beta_\Lambda (x-y)$.
We wish to determine these by demanding that the energy-momentum
tensor have the correct cutoff dependence (\ref{sec4-ERG-comp})
explained in Sec.~\ref{sec-comp}.

Before starting we must ask if it is enough to know the $\Lambda$
dependence of $\alpha_\Lambda$ and $\beta_\Lambda (x-y)$ to determine
them completely.  We start with $\alpha_\Lambda$.  It is a constant
with the mass dimension $D$, the same as the EM tensor
$\Theta_{\mu\nu} (x)$.  Its $\Lambda$ independent part must be
determined by the two parameters $f_1, f_2$ of the theory.  Since the
mass dimension of $f_1$ is $D-2$ and that of $f_2$ is $D-4$, the only
possibility is $\frac{f_1^2}{f_2}$, which diverges at $f_2=0$.  Hence,
we expect that the $\Lambda$ dependence determines $\alpha_\Lambda$
uniquely.  We next consider $\beta_\Lambda (x-y)$, which has the mass
dimension $2D-4$.  The only $\Lambda$ independent possibility is
$f_2 \delta (x-y)$, which would give a term proportional to
\begin{equation}
  f_2 \left(- \delta_{\mu\nu} \partial^2 + \partial_\mu \partial_\nu 
  \right) \sigma (x) \label{sec5-ambiguity}
\end{equation}
to $\Theta_{\mu\nu} (x)$.  We thus conclude that the $\Lambda$
dependence determines $\beta_\Lambda (x-y)$ only up to an additive
term proportional to (\ref{sec5-ambiguity}).  Thus, the Ward identity
(\ref{sec2-Ward-to-solve}) and the symmetry (\ref{sec2-EM-symmetry})
leaves $\Theta_{\mu\nu} (x)$ ambiguous up to a constant multiple of
(\ref{sec5-ambiguity}).

We now determine the $\Lambda$ dependence of both $\alpha_\Lambda$ and
$\beta_\Lambda (x-y)$ by demanding that the trace
$\Theta (x) \equiv \Theta_{\mu\mu} (x)$ have the correct cutoff
dependence.  Eq.~(\ref{sec3-EM}) gives the trace as
\begin{align}
&\frac{1}{N}  \Theta (x)
  = -  D \frac{1}{2} f_2 \sigma (x)^2 \nt\\
  &\quad - (D-2) \frac{1}{2N}
    \partial_\mu \Phi^I (x) \partial_\mu \Phi^I(x)
 + \alpha_{\Theta \Lambda} + \int_y b_{\Theta \Lambda} (x-y)
    \sigma (y)\\
  &\quad + \frac{1}{2} (D-2) \lim_{y \to x}
    \frac{\partial^2}{\partial x_\mu \partial y_\mu}
    \left( \G_\Lambda (x,y) [\sigma] - h_\Lambda (x-y) - \int_z
    h_\Lambda (x-z) \sigma (z) h_\Lambda (z-y) \right)\,,\nt
\end{align}
where we have defined
\begin{subequations}
\begin{align}
  a_{\Theta, \Lambda}
  &\equiv D \alpha_\Lambda,\label{sec5-EM-a}\\
  b_{\Theta,\Lambda} (x-y)
  &\equiv  D A_\Lambda (x-y) - (D-1) \partial^2 \beta_\Lambda (x-y)\,.
\label{sec5-EM-b}
\end{align}
\end{subequations}
The second part of the trace, starting with
$\partial_\mu \Phi^I \partial_\mu \Phi^I$, should belong to the class
of composite operators in Sec.~\ref{sec4-subsec-class}.  Since
\begin{align}
  &\lim_{y\to x} \frac{\partial^2}{\partial x_\mu \partial y_\mu}
    \left( \G_\Lambda (x,y)  - \cdots \right)\nt\\
  &= \frac{1}{2} \left[ \partial_x^2 \lim_{y\to x} \left( \G_\Lambda
    (x,y) \cdots \right) -  \lim_{y\to x} \left(
    \partial_x^2 + \partial_y^2 \right) \left( \G_\Lambda (x,y) -
    \cdots \right) \right]\,,
\end{align}
we obtain
\begin{align}
  \frac{1}{N} \Theta (x)
  &= - D \frac{f_2}{2} \sigma (x)^2   - k f_2 (D-1) \partial^2\sigma
    (x)  - \frac{D-2}{2} \partial^2 \left[ \rho (x)\right]_\Lambda\nt\\
  &\quad - \frac{D-2}{2} \left(\frac{1}{N} \N_\Lambda (x) - 2 f_1 \sigma
    (x) - 2 f_2 \sigma (x)^2 \right),\label{sec5-trace-Nvarphi}
\end{align}
where we have used (\ref{sec4-comp-rho}) and
(\ref{sec4-number-expansion}).  The term proportional to a
dimensionless constant $k$ comes from the additive ambiguity of
$\beta_\Lambda (x-y)$ by $k\,f_2 \delta (x-y)$.

Using (\ref{sec4-rho-a}) and (\ref{sec4-number-ab}), we obtain
\begin{subequations}
  \begin{align}
    a_{\Theta,\Lambda}
    &= \frac{D-2}{2} \int_y \RL (y) h_\Lambda (y)\,,\\
    b_{\Theta,\Lambda} (x)
    &= \frac{D-2}{2} \left[ - 2 c_{1\Lambda} \delta (x)
      + h_\Lambda (x) \int_y \RL (x-y) h_\Lambda (y) - \partial^2
      \frac{1}{2} h_\Lambda (x)^2 \right]\nt\\
    &\quad - k f_2 (D-1) \partial^2 \delta (x)\,.
  \end{align}
\end{subequations}
We then obtain
\begin{subequations}
  \begin{align}
    \alpha_\Lambda
    &= \frac{1}{D} a_{\Theta,\Lambda} = \frac{D-2}{2D} \int_y \RL (y)
      h_\Lambda (y)\,,\label{sec5-EM-alpha}\\
   - \partial^2  \beta_\Lambda (x)
    &= \frac{1}{D-1} \left( b_{\Theta,\Lambda} (x) - D A_\Lambda
      (x)\right) \nt\\
    &= \frac{D-2}{2(D-1)} \left[ - 2 c_{1\Lambda} \delta (x)
      + h_\Lambda (x) \int_y \RL (x-y) h_\Lambda (y) - \frac{2D}{D-2}
      A_\Lambda (x) \right]\nt\\
&\quad  - \partial^2 \lb \frac{D-2}{2(D-1)} \frac{1}{2}
      h_\Lambda (x)^2+  k f_2  \delta (x) \rb\,,\label{sec5-EM-beta}
  \end{align}
\end{subequations}
where $c_{1\Lambda}$ is defined by (\ref{sec2-c1Lambda}), and
$A_\Lambda (x)$ by (\ref{sec3-def-A}).  In Appendix \ref{appendix-B}
we solve (\ref{sec5-EM-beta}) for $\beta_\Lambda (x)$ in the momentum
space.  This completes the construction of the energy-momentum tensor
given by (\ref{sec3-EM}).

We now discuss the trace of the energy-momentum tensor.  From
(\ref{sec5-trace-Nvarphi}), using (\ref{sec4-rho-sigma}), we obtain
\begin{align}
  \frac{1}{N} \Theta (x)
  &= - D f_2 \frac{1}{2} \sigma (x)^2 - k f_2 (D-1) \partial^2 \sigma
    (x) - 
    \frac{D-2}{2} \partial^2 \left( - f_1  - f_2 \sigma (x) \right)\nt\\
  &\quad - \frac{D-2}{2} \left(\frac{1}{N} \N_\Lambda (x) - 2 f_1
    \sigma (x)- 2 f_2 \int_q \sigma (x)^2 \right)\nt\\
  &= - \frac{D-2}{2} \frac{1}{N} \N_\Lambda (x) + (D-2) f_1 \sigma (x)
    + (D-4) f_2 \frac{1}{2} \sigma (x)^2 \nt\\
  &\quad + \left(\frac{D-2}{2} - k (D-1)\right) f_2 \partial^2 \sigma
    (x)\,.    \label{sec5-trace}
\end{align}
Integrating this over space, we obtain
\begin{equation}
\int_x \left(  \Theta (x) + \frac{D-2}{2} \N_\Lambda (x) \right)
  = \left[ (D-2) f_1 \frac{\partial}{\partial f_1} + (D-4) 
    f_2 \frac{\partial}{\partial f_2} \right] N \Gamma_{I\Lambda}
  [\rho],
  \label{sec5-trace-formula}
\end{equation}
where $D-2$ and $D-4$ are the scale dimensions of the parameters $f_1,
f_2$, respectively.  We have thus verified the trace formula that
relates the trace of the energy momentum tensor to the scale
transformation.\cite{Callan:1970ze}

As has been shown in \cite{Polchinski:1987dy}, conformal invariance
amounts to the vanishing of the trace $\Theta (x)$ up to the second
order derivative of a local operator.  Within the context of ERG, we
must find
\begin{equation}
\Theta (x) + \left(\frac{D-2}{2} +
    \gamma\right)\N_\Lambda (x)  = 0
\end{equation}
up to the second order derivative of a composite operator, where
$\gamma$ is the anomalous dimension of the scalar field.  This was
first shown in \cite{Delamotte:2015aaa}, and subsequently discussed in
\cite{Rosten:2014oja} and \cite{Sonoda:2015pva}.  In large $N$ the
anomalous dimension vanishes $\gamma=0$, and at $f_1 = f_2 = 0$,
(\ref{sec5-trace}) gives
\begin{equation}
  \Theta (x) + \frac{D-2}{2} \N_\Lambda (x)
  = 0\,.
\end{equation}
Hence, the theory at $f_1=f_2=$ has conformal invariance.

Equation (\ref{sec5-trace}) for the trace of the EM tensor is
appropriate in the neighborhood of the critical point $f_1=f_2=0$.  In
the neighborhood of the massless Gaussian theory
$f_1=0, f_2 = +\infty$, it is more convenient to rewrite
(\ref{sec5-trace-formula}) using the scalar field $\rho$.  The
squared mass parameter $m_\Lambda^2$, which can be negative, is given by
\begin{equation}
  f_1 - f_2 m_\Lambda^2 +  \frac{1}{2} \int_p \left(
    \frac{1}{p^2+m_\Lambda^2 + R_\Lambda (p)} - \frac{1}{p^2}\right) = 0.
\end{equation}
Since
\begin{equation}
  \Gamma_{I\Lambda} [\rho] \overset{f_2 \to
    +\infty}{\longrightarrow}
\int_x \left(  - m_\Lambda^2 \rho (x) - \frac{1}{2 f_2} \rho (x)^2 \right)
\end{equation}
up to an additive constant if we ignore terms suppressed by
$\frac{1}{f_2^2}$ and higher, we find
\begin{equation}
  \sigma (x) =  \frac{\delta \Gamma_{I\Lambda} [\rho]}{\delta
    \rho (x)} \overset{f_2 \to +\infty}{\longrightarrow}
-  m_\Lambda^2  - \lambda \rho (x),
\end{equation}
where $\lambda \equiv \frac{1}{f_2}$.  Hence,
(\ref{sec5-trace-formula}) gives
\begin{align}
  &  \Theta_\Lambda (x) + \frac{D-2}{2} \N_\Lambda (x)
    \overset{f_2 \to +\infty}{\longrightarrow}
    N \left[  - (D-2)  f_1  \left(
    m_\Lambda^2 +\lambda \rho (x) \right)\right.\nt\\
  &\left.\quad + (D-4) \frac{1}{2 \lambda} \left( m_\Lambda^2  +
    \lambda \rho (x) \right)\left( m_\Lambda^2 
    + \lambda \rho (x) \right) + \left( - \frac{D-2}{2} +
    k(D-1)\right) \partial^2 \rho (x)\right]\nt\\
  &= N \left[ - D \frac{1}{2 \lambda} m_\Lambda^4 
    - 2 m_\Lambda^2 \rho (x)  - (4-D) \frac{\lambda}{2} \rho (x)^2  + \left( - \frac{D-2}{2} +
    k(D-1)\right) \partial^2 \rho (x) \right]\,,
\end{align}
which is a more familiar expression for the trace.  We note that the
scale dimensions of
$\frac{1}{\lambda} m_\Lambda^4, m_\Lambda^2, \lambda$ are respectively
given by $D, 2, 4-D$.

\section{The $\Lambda \to 0+$ limit of $\Theta_{\mu\nu}$\label{sec-zerolimit}}

As we have explained in Sec.~\ref{sec-comp}, a cutoff dependent
composite operator becomes the 1PI effective action with the insertion
of a single composite operator in the vanishing cutoff limit.  Let us
compute the limit
\[
  \lim_{\Lambda \to 0+} \Theta_{\mu\nu} (x)\,.
\]

We find it easier to compute in the momentum space where
\begin{subequations}
\begin{align}
  R_\Lambda (p)
  &= \Lambda^2 R (p/\Lambda)\,,\\
  \tilde{h}_\Lambda (p)
  &= \int_x e^{- i p x} h_\Lambda (x) = \frac{1}{p^2 + R_\Lambda
    (p)} = \frac{1}{\Lambda^2} \frac{1}{\frac{p^2}{\Lambda^2} + R
    (p/\Lambda)} = \frac{1}{\Lambda^2} \tilde{h} (p/\Lambda)
    \,,\\
  \tilde{f}_\Lambda (p)
  &= \int_x e^{- i p x} f_\Lambda (x) = \frac{\Lambda \partial_\Lambda
    R_\Lambda (p)}{\left( p^2 + R_\Lambda (p)\right)^2}
    = \frac{1}{\Lambda^2} \frac{(2 - p \cdot \partial_p) R
    (p/\Lambda)}{ \left(\frac{p^2}{\Lambda^2} + R (p/\Lambda)\right)^2}
     = \frac{1}{\Lambda^2} \tilde{f} (p/\Lambda)\,.
\end{align}
\end{subequations}
From (\ref{sec3-EM}) we find
\begin{align}
&  \tilde{\Theta}_{\mu\nu} (p) \equiv \int_x e^{- i p x} \Theta_{\mu\nu} (x)
  = - N \delta_{\mu\nu} f_2 \frac{1}{2} \int_q \tilde{\sigma} (p+q)
    \tilde{\sigma} (-q)\nt\\
   &\quad + N \left[ \delta_{\mu\nu} \alpha_\Lambda \delta (p)
    + \lb \delta_{\mu\nu} \tilde{A}_\Lambda (p) + \left(p^2
    \delta_{\mu\nu} - p_\mu p_\nu\right) \tilde{\beta}_\Lambda (p)\rb
    \tilde{\sigma} (p) \right] \nt\\
  &\quad +  \frac{1}{2} \int_q \lb
    - \delta_{\mu\nu} q (q+p) + q_\mu (q+p)_\nu + q_\nu (q+p)_\mu
    \rb\nt\\
    &\qquad \times \left[ \tilde{\Phi}^I (-q) \tilde{\Phi}^I (q+p)
    + N  \lb \tilde{\G}_{\Lambda; q,-q-p} [\sigma] - \tilde{h}_\Lambda
    (q) \delta (p) - \tilde{h}_\Lambda (q) \tilde{\sigma} (p)
    \tilde{h}_\Lambda (q+p) \rb  \right]\,,
\end{align}
where
\begin{align}
  &\tilde{\G}_{\Lambda; q,-q-p} [\sigma]
  \equiv \int_{x,y} e^{i q x - i(q+p) y} \G_\Lambda (x,y)
    [\sigma]\nt\\
  &= \tilde{h}_\Lambda (q) \delta (p) + \tilde{h}_\Lambda (q) \Bigg[
    \tilde{\sigma} (p) +  \int_{p_1, p_2} \delta (p_1+p_2-p) \tilde{\sigma} (p_1)
    \tilde{h}_\Lambda (q+p_1) \tilde{\sigma} (p_2)\nt\\
  &\quad + \int_{p_1, p_2, p_3} \delta (p_1+p_2+p_3-p) \tilde{\sigma}
    (p_1) \tilde{h}_\Lambda (q+p_1) \tilde{\sigma} (p_2)
    \tilde{h}_\Lambda (q+p_1+p_2) \tilde{\sigma} (p_3)
+ \cdots \Bigg] \tilde{h}_\Lambda (q+p)\,.
\end{align}

$\alpha_\Lambda$, given by (\ref{sec5-EM-alpha}), vanishes in the limit:
\begin{equation}
  \alpha_\Lambda = \frac{D-2}{2D} \int_q R_\Lambda (q)
  \tilde{h}_\Lambda (q) = \frac{D-2}{2D} \Lambda^D \int_q R (q)
  \tilde{h} (q) \overset{\Lambda \to 0+}{\longrightarrow} 0\,.
\end{equation}
Similarly, (\ref{sec3-def-A}) gives
\begin{align}
  p_\nu \tilde{A}_\Lambda (p)
  &\equiv \int_x e^{- i p x} \frac{1}{i} \partial_\nu A_\Lambda
    (x)\nt\\
  &= \int_q (p+q)_\nu R_\Lambda (q) \tilde{h}_\Lambda (q)
    \tilde{h}_\Lambda (q+p)
  = \Lambda^{D-1} \int_q \left(q + \frac{p}{\Lambda}\right)_\nu R(q)
    \tilde{h}(q) \tilde{h}\left( q + \frac{p}{\Lambda}\right)\nt\\
  &\overset{\Lambda \to 0+}{\longrightarrow}
    p_\nu \frac{\Lambda^D}{p^2} \int_q R (q) \tilde{h} (q) \longrightarrow 0\,.
\end{align}
(\ref{sec5-EM-beta}) gives
\begin{align}
  p^2 \tilde{\beta}_\Lambda (p) 
  &\equiv \int_x e^{- i p x} (-\partial^2) \beta_\Lambda (x)\nt\\
  &= \frac{D-2}{2(D-1)} \left[
    - 2 c_{1\Lambda} + \int_q R_\Lambda (q) \tilde{h}_\Lambda (q)
    \tilde{h}_\Lambda (q+p) - \frac{2D}{D-2} \tilde{A}_\Lambda (p)
    \right]\nt\\
  &\quad + p^2 \lb \frac{D-2}{4(D-1)} \int_q \tilde{h}_\Lambda (q)
    \tilde{h}_\Lambda (q+p) + k f_2 \rb\,,
\end{align}
where (\ref{sec2-c1Lambda}) gives
\begin{equation}
  2 c_{1\Lambda} = \int_q \left(\tilde{h}_\Lambda (q) -   \frac{1}{q^2}\right)
  = \Lambda^{D-2} \int_q \left(\tilde{h} (q)- \frac{1}{q^2}\right)
  \overset{\Lambda \to 0+}{\longrightarrow} 0\,,
\end{equation}
and
\begin{align}
  \int_q R_\Lambda (q) \tilde{h}_\Lambda (q) \tilde{h}_\Lambda (q+p)
  = \Lambda^{D-2} \int_q R(q) \tilde{h} (q) \tilde{h} \left(q +
  \frac{p}{\Lambda}\right)
  \overset{\Lambda \to 0+}{\longrightarrow} \Lambda^D \frac{1}{p^2}
  \int_q R(q) \tilde{h}(q) \longrightarrow 0\,.
\end{align}
Hence, we obtain
\begin{equation}
\lim_{\Lambda \to 0+}  \tilde{\beta}_\Lambda (p) = \frac{D-2}{4(D-1)}
\int_q \frac{1}{q^2 (q+p)^2} + k f_2\,.
\end{equation}

We thus obtain the effective action with the insertion of a single EM
tensor as
\begin{align}
  &\lim_{\Lambda \to 0+} \tilde{\Theta}_{\mu\nu} (p)
  = - N \delta_{\mu\nu} f_2 \frac{1}{2} \int_q \tilde{\sigma} (-q)
  \tilde{\sigma} (q+p) \nt\\
  &\qquad + N \left(p^2 \delta_{\mu\nu} - p_\mu
    p_\nu\right) \left( \frac{D-2}{4(D-1)}
    \int_q \frac{1}{q^2 (q+p)^2} + k f_2\right) \tilde{\sigma} (p)\nt\\
  &\quad + \frac{1}{2} N \int_q \lb - \delta_{\mu\nu} q(q+p) + q_\mu
    (q+p)_\nu + q_\nu (q+p)_\mu \rb \Bigg[ \frac{1}{N} \tilde{\Phi}^I (-q)
    \tilde{\Phi}^I (q+p) \nt\\
  &\qquad +
    \frac{1}{q^2} \lb \int_{p_1, p_2} \delta (p_1+p_2-p)
    \tilde{\sigma} (p_1) \frac{1}{(q+p_1)^2} \tilde{\sigma} (p_2)\right.\nt\\
&\left. +  \int_{p_1, p_2, p_3} \delta
    (p_1+p_2+p_3-p)  \tilde{\sigma} (p_1) \frac{1}{(q+p_1)^2} \tilde{\sigma} (p_2)
    \frac{1}{(q+p_1+p_2)^2} \tilde{\sigma} (p_3) + \cdots \rb
  \frac{1}{(q+p)^2} \Bigg]\,,
    \label{sec6-result}
\end{align}
where $\tilde{\sigma} (p)$ is given by
\begin{equation}
  \tilde{\sigma} (p) \equiv \frac{\delta \Gamma_{\eff, I}
    [\rho]}{\delta \tilde{\rho} (-p)}
\end{equation}
and
\begin{equation}
  \Gamma_{\eff, I} [\rho]  \equiv \lim_{\Lambda \to 0+} \Gamma_{I
    \Lambda} [\rho]\,.
\end{equation}

\section{Conclusions\label{sec-conclusions}}

In this paper we have applied the exact renormalization group
formalism to construct the energy-momentum (EM) tensor in the large
$N$ limit of the O($N$) linear sigma model in $D$ dimensions, where
$2 < D < 4$.  The EM tensor is a functional of scalar fields, and its
cutoff dependence is given by the exact renormalization group
equation.  By taking the momentum cutoff to zero, we obtain the
effective action with a single insertion of the energy-momentum
tensor.

Our result for the cutoff dependent EM tensor is given by
(\ref{sec3-EM}) of Sec.~\ref{sec-EM}, where the coefficient
$A_\Lambda (x)$ is given by (\ref{sec3-def-A}), and $\alpha_\Lambda$
and $\beta_\Lambda (x)$ are given respectively by
(\ref{sec5-EM-alpha}), (\ref{sec5-EM-beta}).  The field $\sigma (x)$
is defined by (\ref{sec2-sigma}).  The EM tensor has been constructed
to satisfy the Ward identity (\ref{sec2-Ward-to-solve}) for the
translation and rotation invariance.  In Sec.~\ref{sec-zerolimit} we
take the zero cutoff limit to obtain (\ref{sec6-result}).

In general, given a Wilson action, we should be able to write down the
energy-momentum tensor explicitly.  We hope we have presented such an
example for an interacting theory even though we have needed the help
of large $N$ approximations.  We have verified that our
energy-momentum tensor satisfies the expected trace formula.  We have
also confirmed that the critical O$(N)$ model is conformally invariant
in the large $N$ limit (see \cite{DePolsi:2019owi} for a different
approach valid for any finite $N$).

Now that the EM tensor is constructed explicitly within the ERG
formalism, we should be able to compute also the short distance
singularities in the products of the tensor.  
This is a task left for the future; we refer the reader to
\cite{Pagani:2017tdr,Pagani:2020ejb} for some explicit examples of
calculations of short distance singularities of operator products in
the ERG framework.
Another issue left for a future work is to find the
relation between the EM tensor defined in this paper and the EM tensor
defined by coupling the system to an external metric and possibly an
additional background metric.
In this case we expect that the Ward identity associated with the
shift of the background metric also plays a role;  it would be
interesting to study this along the lines of \cite{Pagani:2024lcn}.
This approach could have some merit also for the study of the role of
these Ward identities within the asymptotic safety scenario for
quantum gravity \cite{Reuter:2019byg,Percacci:2017fkn}.

\newpage

\appendix

\section{Derivation of (\ref{sec4-ERG-OL})\label{appendix-A}}

A cutoff dependent composite operator $\Op_\Lambda [J]$ is defined by
(\ref{sec4-OL-J}).  Eq.~(\ref{sec4-ERG-OL}) gives the cutoff
dependence of $\Op_\Lambda [\Phi] = \Op_\Lambda [J]$ regarded as a
functional of $\Phi = \frac{\delta W_\Lambda [J]}{\delta J}$ instead
of $J$.  It is the purpose of this Appendix to derive
(\ref{sec4-ERG-OL}).

First, the cutoff dependence of $\Op_\Lambda [J]$ as a
functional of $J$ is obtained, from (\ref{sec4-OL-J}), as
\begin{equation}
  - \Lambda \partial_\Lambda \Op_\Lambda [J]
  = \frac{1}{2} \int_{x,y} \Lambda \partial_\Lambda \RL (x-y) \lb
  \frac{\delta^2 \Op_\Lambda [J]}{\delta J(x) \delta J(y)} + 2
  \frac{\delta \Op_\Lambda [J]}{\delta J(x)} \frac{\delta W_\Lambda
    [J]}{\delta J(y)} \rb\,.
\end{equation}
We then obtain
\begin{align}
  - \Lambda \partial_\Lambda \Op_\Lambda [\Phi]
  &= - \Lambda \partial_\Lambda \Op_\Lambda [J] + \int_x \frac{\delta
    \Op_\Lambda [J]}{\delta J(x)} (- \Lambda \partial_\Lambda)
  J(x)\Big|_{\Phi\,\textrm{fixed}}\nt\\
  &=  - \Lambda \partial_\Lambda \Op_\Lambda [J] + \int_x \frac{\delta
    \Op_\Lambda [J]}{\delta J(x)} (- \Lambda \partial_\Lambda)
    \left( \int_y \RL (x-y) \Phi (y) - \frac{\delta \Gamma_\Lambda
    [\Phi]}{\delta \Phi (x)} \right)\,,
\end{align}
where (\ref{sec2-JPhi}) is used.  Using
\begin{align}
  \frac{\delta \Op_\Lambda [J]}{\delta J(x)}
  &= \int_y \frac{\delta \Phi (y)}{\delta J(x)} \frac{\delta
    \Op_\Lambda [\Phi]}{\delta \Phi (y)}
= \int_y \G_\Lambda (x,y) [\Phi] \frac{\delta \Op_\Lambda
    [\Phi]}{\delta \Phi (y)}\,,\\
  \frac{\delta^2 \Op_\Lambda [J]}{\delta J(x) \delta J(y)}
  &= \int_v \G_\Lambda (y,v) [\Phi] \frac{\delta}{\delta \Phi (v)}
    \int_u \G_\Lambda (x,u) [\Phi] \frac{\delta \Op_\Lambda
    [\Phi]}{\delta \Phi (u)}\nt\\
  &= \int_{u, v} \G_\Lambda (y,v) [\Phi] \left(
    \frac{\delta \G_\Lambda (x,u)}{\delta \Phi (v)} \frac{\delta \Op_\Lambda
    [\Phi]}{\delta \Phi (u)} + \G_\Lambda (x,u) [\Phi] \frac{\delta^2
    \Op_\Lambda [\Phi]}{\delta \Phi (u) \delta \Phi (v)} \right)\,,\\
  - \Lambda \partial_\Lambda \frac{\delta \Gamma_\Lambda
    [\Phi]}{\delta \Phi (x)}
 &= \frac{\delta}{\delta \Phi (x)} \frac{1}{2} \int_{u,v} \Lambda
  \partial_\Lambda \RL (u-v) \, \G_\Lambda (u,v) [\Phi]\nt\\
&= \frac{1}{2} \int_{u,v} \Lambda
  \partial_\Lambda \RL (u-v) \, \frac{\delta G_\Lambda (u,v)
  [\Phi]}{\delta \Phi (x)}\,,
\end{align}
we obtain
\begin{equation}
  - \Lambda \partial_\Lambda \Op_\Lambda [\Phi]
  = \frac{1}{2} \int_{x,y} \Lambda \partial_\Lambda \RL (x-y)
  \int_{u,v} \G_\Lambda (x,u) \G_\Lambda (y,v) \frac{\delta^2
    \Op_\Lambda [\Phi]}{\delta \Phi (u) \delta \Phi (v)}
\end{equation}
which is (\ref{sec4-ERG-OL}).  For the linear sigma model, this
generalizes to
\begin{equation}
  - \Lambda \partial_\Lambda \Op_\Lambda [\Phi]
  = \frac{1}{2} \int_{x,y} \Lambda \partial_\Lambda \RL (x-y)
  \int_{u,v} \G^{IJ}_\Lambda (x,u) \G^{IK}_\Lambda (y,v) \frac{\delta^2
    \Op_\Lambda [\Phi]}{\delta \Phi^J (u) \delta \Phi^K (v)}\,.
\end{equation}

\section{Solution to (\ref{sec5-EM-beta}) \label{appendix-B}}

In this Appendix we solve (\ref{sec5-EM-beta}) in the momentum space,
where the equation becomes
\begin{align}
  p^2 \left(\tilde{\beta}_\Lambda (p) - k f_2 \right)
  &= \frac{1}{D-1} \left( \tilde{b}_{\Theta, \Lambda} (p) - D
    \tilde{A}_\Lambda (p) \right)\,\nt\\
  &= \frac{1}{D-1} \frac{D-2}{2} \left[ \int_q \left(R_\Lambda (q)
    \tilde{h}_\Lambda (q) \tilde{h}_\Lambda (q+p) - h_\Lambda (q) +
    \frac{1}{q^2}\right) \right.\nt\\
  &\left.\quad- \frac{2D}{D-2} \tilde{A}_\Lambda (p) + \frac{1}{2}
    p^2 \int_q \tilde{h}_\Lambda (q) \tilde{h}_\Lambda (q+p) \right]\,.
\end{align}
This is solved by
\begin{align}
&  \tilde{\beta}_\Lambda (p)
  = k f_2 + \frac{1}{D-1}\frac{D-2}{2} \frac{1}{2} \int_q
    \tilde{h}_\Lambda (q) \tilde{h}_\Lambda   (q+p)\nt\\
  &\quad +  \frac{1}{D-1}\frac{D-2}{2} \frac{1}{p^2}
    \left[
    \int_q \left(R_\Lambda (q) \tilde{h}_\Lambda (q) \tilde{h}_\Lambda (q+p) -
    \tilde{h}_\Lambda (q) + \frac{1}{q^2}\right)
    - \frac{2D}{D-2} \tilde{A}_\Lambda (p)\right]\,.
\end{align}

For this solution to make sense, we must show that the coefficient
function $\tilde{\beta}_\Lambda (p)$ defined by (\ref{sec5-EM-beta}) is finite
at $p=0$.  We need to show
\begin{equation}
  \frac{2D}{D-2} \tilde{A}_\Lambda (0) = \int_q \left(R_\Lambda (q) \tilde{h}_\Lambda
    (q)^2 - \tilde{h}_\Lambda (q) + \frac{1}{q^2}\right)\,,\label{app-equality}
\end{equation}
where $\tilde{A}_\Lambda (p)$ is determined by (\ref{sec3-def-A}) as
\begin{equation}
  p_\nu \tilde{A}_\Lambda (p) = \int_q (p+q)_\nu R_\Lambda (q)
  \tilde{h}_\Lambda (q) \tilde{h}_\Lambda (q+p)\,.
\end{equation}

Since
\begin{align}
p_\nu \tilde{A}_\Lambda (p)  &=   \int_q (q+p)_\nu R_\Lambda (q) \tilde{h}_\Lambda (q)
                       \tilde{h}_\Lambda (q+p)\nt \\
  &\overset{p \to 0}{\longrightarrow}
     p_\nu \int_q R_\Lambda (q) \tilde{h}_\Lambda (q)^2 +  \int_q q_\nu
    R_\Lambda (q) \tilde{h}_\Lambda (q) p_\mu \frac{\partial}{\partial q_\mu}
    \tilde{h}_\Lambda (q)\nt\\
  &=  p_\nu \left( \int_q R_\Lambda (q) \tilde{h}_\Lambda (q)^2 + \frac{1}{D}
    \int_q R_\Lambda (q) \tilde{h}_\Lambda (q) q \cdot \partial_q \tilde{h}_\Lambda
    (q) \right),
\end{align}
we obtain
\begin{align}
  \tilde{A}_\nu (0)
  &=  \int_q R_\Lambda (q) \tilde{h}_\Lambda (q)^2 + \frac{1}{2D}
    \int_q R_\Lambda (q) q \cdot \partial_q \tilde{h}_\Lambda (q)^2\nt\\
  &=  \int_q R_\Lambda \tilde{h}_\Lambda^2 + \frac{1}{2D} \int_q
    \left(-D-q\cdot\partial_q\right) R_\Lambda (q) \cdot \tilde{h}_\Lambda
    (q)^2\nt\\
  &=  \int_q R_\Lambda \tilde{h}_\Lambda^2 + \frac{1}{2D} \int_q
    \left(2-q\cdot\partial_q - D-2\right) R_\Lambda (q) \cdot \tilde{h}_\Lambda
    (q)^2\nt\\
  &=  \int_q R_\Lambda \tilde{h}_\Lambda^2 + \frac{1}{2D} \int_q \left(
    \tilde{f}_\Lambda (q) - (D+2) R_\Lambda (q) \tilde{h}_\Lambda (q)^2 \right)\nt\\
  &= \frac{D-2}{2D}  \int_q R_\Lambda (q) \tilde{h}_\Lambda (q)^2 + \frac{1}{2D} \int_q
    \tilde{f}_\Lambda (q).\label{app-equality-prep}
\end{align}
Using
\begin{equation}
  \frac{1}{D} \int_q \tilde{f}_\Lambda (q) = \frac{D-2}{D} \int_q
  \left(\frac{1}{q^2}-\tilde{h}_\Lambda (q)\right)\,,
\end{equation}
we obtain the desired equality (\ref{app-equality}).


\bibliography{paper}

\end{document}